\documentclass[prd,aps,twocolumn,a4paper,nofootinbib]{revtex4} 
\usepackage{graphicx}
\usepackage[breaklinks=true,colorlinks=true,linkcolor=blue,urlcolor=blue,citecolor=blue]{hyperref}

\usepackage{color}

\usepackage[utf8x]{inputenc}
\usepackage{subeqn}
\usepackage{wasysym}
\usepackage{tikz}
\usepackage{array}
\usepackage{verbatim}
\usepackage{orcidlink}
\usepackage{xspace}

\usepackage{mathrsfs}
\DeclareSymbolFontAlphabet{\mathrsfs}{rsfs}
\newcommand{\scri}{\mathrsfs{I}}
\newcommand{\scrip}{$\scri^+$}

\newcommand{\aconf}{\bar\Omega}

\newcommand{\alexthesis}{Vano-Vinuales:2015lhj}

\newcommand{\Kc}{K_{\textrm{\tiny CMC}}}
\newcommand{\Cc}{C_{\textrm{\tiny CMC}}}

\newcommand{\tort}{\tilde r_*}
\newcommand{\trum}{_{\textrm{\footnotesize trum}}}
\newcommand{\eref}[1]{(\ref{#1})}

\newcommand{\polyK}{\kappa}
\newcommand{\enerdens}{e}

\newcommand{\rtilde}{{\tilde r}}
\newcommand{\ttilde}{{\tilde t}}
\newcommand{\rtsurf}{{\rtilde_{\textrm{\footnotesize surf}}}}
\newcommand{\rbar}{{\bar r}}
\newcommand{\rsurf}{{r_{\textrm{\footnotesize surf}}}}
\newcommand{\rbsurf}{{\rbar_{\textrm{\footnotesize surf}}}}
\newcommand{\mns}{m_{\textrm{\tiny NS}}}
\newcommand{\mbh}{m_{\textrm{\tiny BH}}}
\newcommand{\msm}{M_{\textrm{\tiny MS}}}
\newcommand{\sns}{\textrm{\tiny NS}}
\newcommand{\sbh}{\textrm{\tiny BH}}
\newcommand{\rbhor}{\bar r_{\sbh}}
\newcommand{\rthor}{\tilde r_{\sbh}}
\newcommand{\Aa}{f}
\newcommand{\rttort}{\tilde r_{*}}
\newcommand{\B}{\tilde g_{\rtilde\rtilde}}

\newcommand{\sref}[1]{section~\ref{#1}}
\newcommand{\ssref}[1]{subsection~\ref{#1}}
\newcommand{\aref}[1]{appendix~\ref{#1}}
\newcommand{\fref}[1]{figure~\ref{#1}}
\newcommand{\Fref}[1]{Figure~\ref{#1}}

\begin{document}

\title[]{Hyperboloidal neutron star and black hole initial data in spherical symmetry}  

\author{Alex Vañó-Viñuales$^{1,2,*}\,\orcidlink{0000-0002-8589-006X}$} 
\affiliation{$^1$Departament de Física, Universitat de les Illes Balears, IAC3, Carretera Valldemossa km 7.5, E-07122 Palma, Spain\\
$^2$Centro de Astrof\'{\i}sica e Gravita\c c\~ao - CENTRA, Departamento de F\'isica, Instituto Superior T\'ecnico IST, Universidade de Lisboa UL, Avenida Rovisco Pais 1, 1049-001 Lisboa, Portugal}
\email{alex.vano@uib.es}

\begin{abstract}

The focus of this work is on the construction of initial data including a neutron star on a hyperboloidal slice. As simplest scenario for this first step, spherical symmetry is considered together with a polytropic-like equation of state for the neutron star. Constraint-satisfying hyperboloidal initial data are obtained for a single neutron star and for a combination of neutron star with a black hole in its center. To the author's best knowledge this is the first time that full hyperboloidal slices of a neutron star spacetime are constructed. The obtained initial data are suitable for evolutions of the Einstein and relativistic hydrodynamic equations on hyperboloidal slices. 

\end{abstract}

\maketitle

\section{Introduction}

Numerical relativity simulations of compact binary systems including black holes (BHs) and neutron stars (NSs) are of utmost importance for the theoretical modelling of the gravitational wave signals measured by gravitational wave interferometers. These are idealized~\cite{Leaver1986} as being located at future null infinity (\scrip, see diagrams in figures~\ref{penns} and~\ref{penboth}), the endpoints of future-directed null geodesics. Future null infinity is also the only location in spacetime where gravitational radiation is unambiguously defined, so that including it in numerical simulations is crucial for avoiding systematic errors associated with waveform extraction. 

Besides Cauchy-characteristic evolution~\cite{Bishop:1996gt,Taylor:2013zia,Moxon:2021gbv} and matching~\cite{BishopCCM,Bishop:1998ukk,Ma:2024hzq}, the hyperboloidal method~\cite{friedrich1983,Zenginoglu:2008pw} is a suitable and promising approach to numerically reach \scrip, despite still being under development for three-dimensional simulations. It relies on the evolution on hyperboloidal slices~\cite{Zenginoglu:2007jw,Zenginoglu:2024bzs}, which are smooth and spacelike, and span all the way to null infinity. Simulations on hyperboloidal slices in the last decades have been carried out with constrained evolutions~\cite{Rinne:2009qx,Rinne:2013qc,Morales:2016rgt}, in a modern setup~\cite{Frauendiener:2023ltp,Frauendiener:2025xcj} with the Conformal Einstein Field Equations~\cite{friedrich1983}, with conformal compactification~\cite{Vano-Vinuales:2014koa,Vano-Vinuales:2017qij,Vano-Vinuales:2024tat} and with the dual-frame approach~\cite{Peterson:2024bxk}. The last two approaches have delivered results in spherical symmetry so far, with extensions to three spatial dimensions currently underway. 

Initial data including BHs in spherical and in axial symmetry were presented in \cite{Schneemann}, while \cite{Schinkel:2013tka,Schinkel:2013zm} dealt with axisymmetric asymptotic constant-mean-curvature (CMC) slices for Kerr. Binary configurations of boosted and spinning BHs were considered in \cite{Buchman:2009ew,schinkelthesis}. These works treat the BHs via excision~\cite{Seidel:1992vd}, cutting away the spacetime inside of the horizon to avoid the singularity. Trumpet constructions~\cite{Hannam:2006xw,Hannam:2006vv,Hannam:2008sg}, which choose a specific slice that only asymptotically reaches the singularity and is suitable for puncture~\cite{1963PhRv..131..471B,Brandt:1997tf,Beig:1994rp} evolutions~\cite{Campanelli:2005dd,Baker:2005vv}, were considered for hyperboloidal evolutions in~\cite{Vano-Vinuales:2023yzs}. 

This work extends the construction of initial data on hyperboloidal slices to NSs spacetimes, preparing them for hyperboloidal evolutions. In this first step, a polytropic equation of state is considered for the NS. The spherically symmetric configuration of a polytropic NS with a BH in its center from~\cite{Richards:2021upu} is also tackled. 
The purpose of~\cite{Richards:2021upu} was to simulate relativistic Bondi accretion~\cite{10.1093/mnras/112.2.195}, a very relevant dynamical scenario in spherical symmetry. A challenging future test for hyperboloidal evolutions will be the same scenario of a NS that is accreted by a parasitic BH in its interior, for which here suitable hyperboloidal initial data are constructed. 
As the aim of this work is to express the given spacetime on a hyperboloidal slice, considering a substantially larger BH than done in~\cite{Richards:2021upu} helps validate the developed method, which is also valid for small BHs. 

The steps to express the desired spherically symmetric configuration in compactified hyperboloidal slices are pedagogically explained and consist basically of the following. Starting from the metric on a Cauchy slice, it is to be represented in terms of an areal radial coordinate. Then a change in the time coordinate will transform it to a hyperboloidal slice (``hyperboloidalization'') using the height function approach~\cite{10.1063/1.1666384,Malec:2003dq,Calabrese:2005rs}, and a compression of the radial coordinate will allow to put \scrip\xspace at finite coordinate distance (``compactification''). 
New in this work is the application of the hyperboloidalization and compactification procedures in subsections \ref{nshyp}, \ref{nscompact}, \ref{bothhyp} and \ref{bothcompact}. 
The presented procedure creates spherically symmetric hyperboloidal initial data ready to be evolved, which is the future next step. 

This paper is organised as follows: \sref{ns} reviews the construction of a polytropic NS and then expresses it on a compactified hyperboloidal slice. The spherically symmetric superposition of NS+BH of \cite{Richards:2021upu} is considered in \sref{nsbh}: first the separate NS and BH solutions (as functions of the isotropic radius) are superposed, second the Hamiltonian constraint is solved, and after that the solution is expressed in terms of the areal radius, required for the posterior hyperboloidalization and compactification. A few concluding remarks and future plans are gathered in \sref{conclu}. Details on the transformation to and from areal radius for BHs, and the creation of conformal Carter-Penrose diagrams are included respectively in appendices \ref{isoareal} and \ref{diagrams}. {Appendix~\ref{inidata} explains how the constructed spacetime slices are set as initial data for the future evolutions.}
Regarding notation, the chosen metric signature is $(-, +, +, +)$ and the fundamental constants are set to $G = c = 1$ (geometrized units) -- see comment on NS parameters in~\ssref{cauchytov}. 
A companion \texttt{Mathematica} notebook covering all steps of the derivation is available in~\cite{nsbhrepo}. 

\section{Initial data for a neutron star}\label{ns} 

The construction of hyperboloidal initial data for a Schwarzschild BH in CMC trumpet form is explained in detail in \cite{Vano-Vinuales:2023yzs}. Here the procedure is adapted for a polytropic-like NS. 

\subsection{Polytropic-like neutron star on Cauchy slice}\label{cauchytov}

For completeness let us first briefly review the construction of NS initial data via the Tolman-Oppenheimer-Volkoff (TOV) equations. 
We start with the spherically symmetric metric ansatz for the metric $\tilde g_{ab}$ of the NS given as the line element 
\begin{equation}\label{metricuncomp}
d\tilde s^2 = -e^{\nu(\rtilde)}d\ttilde^2+e^{\lambda(\rtilde)}d\rtilde^2+\rtilde^2d\sigma^2 ,
\end{equation}
expressed in the usual time $\ttilde$ and uncompactified radial coordinate $\rtilde$, and where $d\sigma^2=d\theta^2+\sin^2\theta\,d\varphi^2$. 

The stress-energy-momentum tensor for a static spherically symmetric perfect fluid is
\begin{equation}
T_{ab} = (\enerdens+P)\tilde u_a\tilde u_b+P\tilde g_{ab}, 
\end{equation}
where $\enerdens(\rtilde)$ is the proper energy density of the fluid, $P(\rtilde)$ its pressure and $\tilde u_a=(e^{\nu(\rtilde)/2},0,0,0)$ the fluid's 4-velocity. 

It is useful to introduce the mass $m(\rtilde)$ contained within a radius $\rtilde$ as measured by the gravitational field felt by a distant observer. It satisfies the relation
\begin{equation}\label{meq}
\frac{dm(\rtilde)}{d\rtilde}\equiv\partial_\rtilde m(\rtilde)= 4 \pi  \rtilde^2 \enerdens(\rtilde) . 
\end{equation}
Substituting $\enerdens(\rtilde)$ in terms of $\partial_\rtilde m(\rtilde)$ from above into the $\ttilde\ttilde$ component of the Einstein equations yields an equation for $\lambda(\rtilde)$, with solution
\begin{equation}\label{lambdaeq}
e^{\lambda(\rtilde)} = \left(1-\frac{2m(\rtilde)}{\rtilde}\right)^{-1}. 
\end{equation}
From now on, either the left or the right expression above will be set for the $\rtilde\rtilde$ component of the NS metric depending on convenience. 

The $\rtilde\rtilde$ component provides an equation for $\nu(\rtilde)$
\begin{equation}\label{nueq}
\partial_\rtilde \nu(\rtilde) = 2\left(\frac{m(\rtilde)}{\rtilde^2}+4 \pi\rtilde P(\rtilde)\right)\left(1-\frac{2  m(\rtilde)}{\rtilde}\right)^{-1},
\end{equation}
and from $\nabla _{\mu }T_{\rtilde}^{\mu } = 0$ we obtain
\begin{equation}\label{p1eq}
{\partial_\rtilde P(\rtilde)}=-{\frac {1}{2}}\left(P(\rtilde)+\enerdens(\rtilde) \right) {\partial_\rtilde \nu(\rtilde)}.
\end{equation}
Substituting \eref{nueq} into the above, or equivalently considering the $\theta\theta$ component of the Einstein equations and substituting \eref{nueq} and its radial derivative yields 
\begin{equation}\label{p2eq}
\partial_\rtilde P = -\left(P+\enerdens \right) \left(\frac{m}{\rtilde^2}+4 \pi\rtilde P\right)\left(1-\frac{2  m}{\rtilde}\right)^{-1} ,
\end{equation}
where the dependence on $\rtilde$ of $P, \enerdens$ and $m$ has not been written explicitly. 

To close the system, the chosen equation of state is a polytropic-like\footnote{The actual polytropic equation relates the pressure to the baryonic density $\rho_0$ (instead of the energy density $\enerdens$) as $P=\polyK\rho_0^{\Gamma}$, with $\enerdens=\rho_0(1+\epsilon)$ and $\epsilon$ the specific internal energy. Here equation~\eref{eos} has been chosen for its simplicity. As the solution is obtained numerically anyway, applying the procedure that follows to a proper polytrope should be straightforward.} equation (called ``$e$-polytrope'' in~\cite{Damour:2009vw}), 
\begin{equation}\label{eos}
P(\rtilde)=\polyK[\enerdens(\rtilde)]^{\Gamma}=\polyK[\enerdens(\rtilde)]^{1+1/n} , 
\end{equation}
which is used to eliminate $P$ in terms of $\enerdens$. The chosen values of the parameters are the common $\Gamma=2$, and $\polyK=1$. The latter was customary used some time ago \cite{Baumgarte:1997eg}, and can be set here without any loss of generality. Equations of state generally break the scale-invariance introduced by the use of geometrized units, so additional choices are required. As $\polyK$ scales masses and radii, it can serve this purpose. For indications on how to convert the results to different values of $\polyK$ see e.g.~\cite{Anderson:2007kz}'s appendix, \cite{Noble:2003xx}'s appendix and~\cite{wald}'s appendix F. 

\begin{figure}[h]
\includegraphics[width=0.95\linewidth]{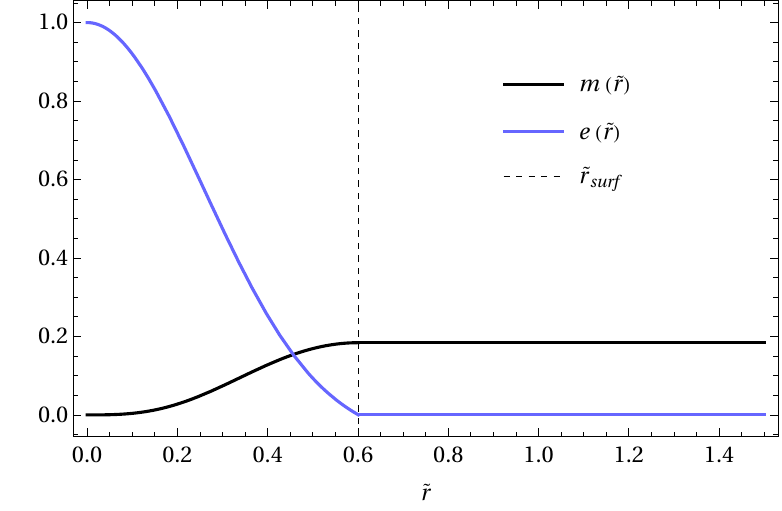}
\vspace{-2ex}
\caption{Solutions $m(\rtilde)$ and $\enerdens(\rtilde)$ of \eref{meq} and \eref{p2eq} with \eref{eos} and $\polyK=1$ and $\Gamma=2$. The vertical dashed line denotes the location of $\rtsurf$, the surface of the NS.}\label{tovsol}
\end{figure}

The numerical resolution of the system starts by setting a value of the energy density at the center of the star,for simplicity taken to be $\enerdens_{\textrm{center}}=1$\footnote{This choice of central density provides a solution that is gravitationally unstable. This is not relevant here, as the NS is only considered as initial data (not evolution) and this work's purpose is to develop a suitable hyperboloidalization procedure, which will be the same for stable or unstable configurations.}. Then equations \eref{meq} and \eref{p2eq}, after substitution of $P$ by \eref{eos}, are solved\footnote{The solutions to most equations in this work were obtained numerically using \texttt{Mathematica}'s function \texttt{NDSolve}.} together from the origin\footnote{The equations are formally singular at the origin, so in practice the integration starts at a very small radius, here $\rtilde=10^{-35}$. Alternatively, the limit at the origin can be derived, but it is a more involved procedure yielding almost equal numerical results.} outwards for $m(\rtilde)$ and $\enerdens(\rtilde)$ until the latter is zero, which corresponds to the surface of the star $\rtsurf$. The results are displayed in~\fref{tovsol}. Outside of the star the spacetime is described by the Schwarzschild metric \eref{metricschw} with a mass equal to 
\begin{equation}
\mns\doteq m(\rtsurf) .
\end{equation}
This allows to set the condition $\nu(\rtsurf) = \log\left(1 - \frac{2\mns}{\rtsurf}\right)$ for $\nu(\rtilde)$ at the star's surface, so that it matches the Schwarzschild value. This condition is used to solve \eref{nueq} from $\rtsurf$ inwards to $\rtilde=0$ for $\nu(\rtilde)$ using the previously obtained solutions of $m(\rtilde)$ and $\enerdens(\rtilde)$. 
Alternatively, $\nu(\rtilde)$ can be calculated using the constructed $P(\rtilde)$ via integral~(3) and evaluation of~(8) in~\cite{Kyutoku:2025zud} (note that there is a factor 2 of difference between the definitions of $\nu$ here and there). 
The final expressions for $m(\rtilde)$, $\enerdens(\rtilde)$ and the $\tilde g_{\ttilde\ttilde} = -e^{\nu(\rtilde)}$ and $\tilde g_{\rtilde\rtilde} = e^{\lambda(\rtilde)}$ metric components are put piecewise together by matching to the outer Schwarzschild expressions with mass $\mns$. 

\subsection{Hyperboloidalization of single neutron star}\label{nshyp}

The metric found above for the polytropic-like NS can be expressed on a hyperboloidal slice using the height function approach \cite{Gentle:2000aq,Malec:2003dq}, where the usual time coordinate $\ttilde$ is transformed to the hyperboloidal time $t$ via 
\begin{equation}\label{ttrafo}
\ttilde = t + h(\rtilde) ,
\end{equation}
where $h(\rtilde)$ is the height function. The height function's effect is to ``raise'' the slices so that they reach \scrip instead of spacelike infinity $i^0$ -- see diagram on the right in~\fref{penns} for a visual representation. One may naively think that, as the height function affects the asymptotic region, it need not have any knowledge of what is contained in the center of the spacetime, so that a height function that works for Minkowski will also work for a NS or a BH. This is not correct: taking into account the mass/energy contained in the system is required for successfully constructing the height function. Otherwise, its asymptotic fall-off behaviour will not allow the slices to intersect \scrip. This matches the discussion around figure 1 in \cite{Peterson:2024bxk}. 

The metric on a hyperboloidal slice within the height function approach, after substituting \eref{ttrafo} in \eref{metricuncomp}, yields
\begin{eqnarray}\label{metricnshyp}
d\tilde s^2 = &&-e^{\nu(\rtilde)}dt^2 -2e^{\nu(\rtilde)}h'(\rtilde)dt\,d\rtilde \nonumber \\ && +\left(e^{\lambda(\rtilde)}-e^{\nu(\rtilde)}\left[h'(\rtilde)\right]^2\right)d\rtilde^2+\rtilde^2d\sigma^2 .
\end{eqnarray}
The quantity to be determined is the boost, the derivative of the height function $h'(\rtilde)=\partial_\rtilde h(\rtilde)$. 
To do this for the TOV solution of the polytropic-like NS obtained above in \ssref{cauchytov}, we follow the derivation in e.g. subsection 3.2.2 in \cite{\alexthesis}, where CMC slices (with constant trace $\tilde K$ of the physical extrinsic curvature) are constructed. The condition is expressed as 
\begin{equation}\label{cmccond}
\tilde{ K} = -\frac{1}{\sqrt{-\tilde g}}\partial_a\left(\sqrt{-\tilde g}\,\tilde n^a\right) \equiv \textrm{constant} , 
\end{equation}
where $\tilde g$ is the determinant of the metric and $\tilde n^a$ is the future pointing unit normal to the constant $\ttilde$ hypersurfaces. In terms of the metric \eref{metricnshyp}, 
\begin{equation}\label{gnns}
-\tilde g = e^\nu e^\lambda\rtilde^4\sin^2\theta , \quad
\tilde n^a = \left(\begin{array}{c}\sqrt{\frac{e^\lambda-e^\nu (h')^2}{e^\nu e^\lambda}}\\\frac{e^\nu h'}{\sqrt{e^\nu e^\lambda(e^\lambda-e^\nu (h')^2)}}\\0\\0\end{array}\right) . 
\end{equation}
 Substituting the above in \eref{cmccond} yields
\begin{equation} \label{hk}
\partial_\rtilde\left(\frac{\rtilde^2e^\nu h'}{\sqrt{e^\lambda-e^\nu(h')^2}}\right) = -\tilde K\rtilde^2\sqrt{e^\lambda e^\nu} , 
\end{equation}
\begin{eqnarray} \label{hkint} 
\frac{e^\nu h'}{\sqrt{e^\lambda-e^\nu(h')^2}} &=& -\frac{1}{\rtilde^2}\left[\int\Kc\rtilde^2\sqrt{e^\lambda e^\nu}d\rtilde +\Cc\right] \nonumber \\
&\equiv& int(\rtilde) ,
\end{eqnarray}
where all quantities are functions of $\rtilde$. $\tilde K$ is taken to be the constant parameter $\Kc$, set to $-3$ from now\footnote{For clearer visualization in the Penrose diagrams in figures~\ref{penns} and~\ref{penboth}.}, and the integration constant $\Cc$ is used to match the interior and exterior solutions. 
Isolating from above, the expression for the boost in term of the shorthand $int(\rtilde)$ is 
\begin{equation}\label{boostexprin}
h'(\rtilde) = \pm int(\rtilde) \sqrt{\frac{e^{\lambda(\rtilde)}}{e^{\nu(\rtilde)}(e^{\nu(\rtilde)}+[int(\rtilde)]^2)}} ,
\end{equation}
where the positive sign is chosen so that $h'(\rtilde)$ is positive asymptotically (for slices reaching \scrip, $\Kc<0$). 

\begin{figure}[h]
\includegraphics[width=0.95\linewidth]{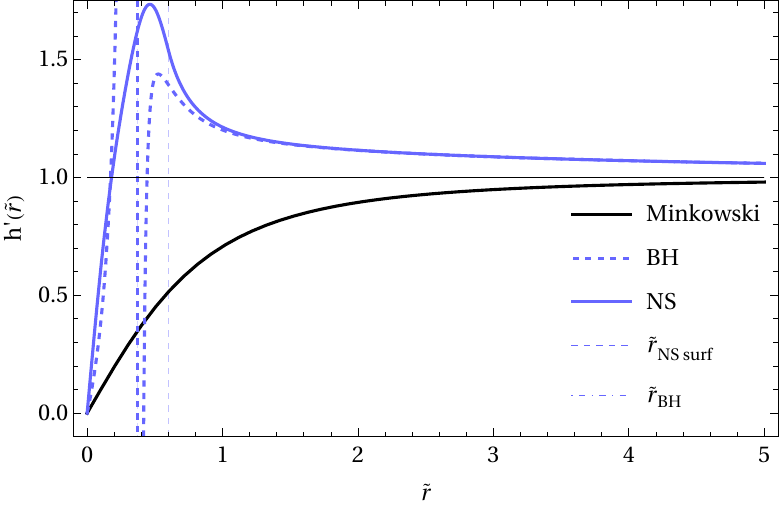}
\\
\includegraphics[width=0.95\linewidth]{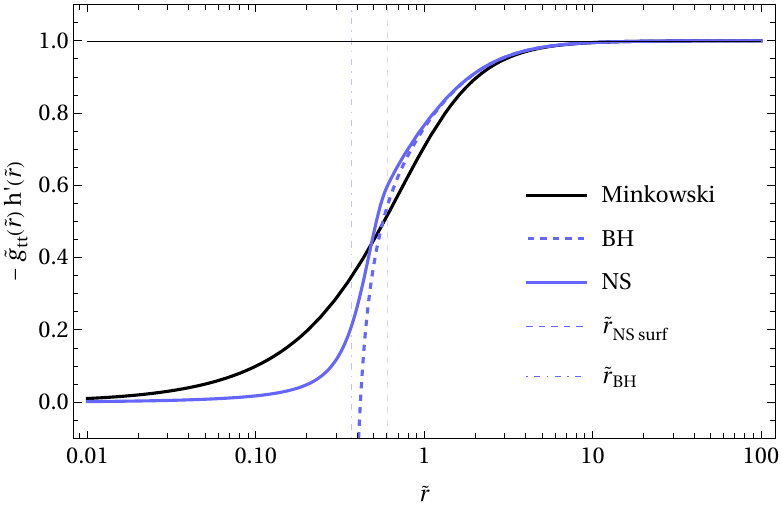}
\vspace{-2ex}
\caption{Boost on top and at the bottom boost multiplied by $-\tilde g_{tt}$, for Minkowski spacetime, for a Schwarzschild BH with mass $\mns$ and for the NS solution found. The dashed line represents the surface of the star and the dot-dashed one, the location of the horizon of a BH with the same mass.}\label{boost}
\end{figure}

\begin{figure}[h]
\includegraphics[width=0.45\linewidth]{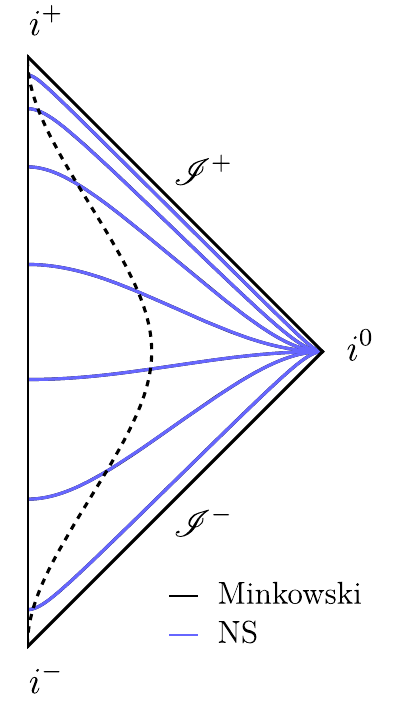}
\includegraphics[width=0.45\linewidth]{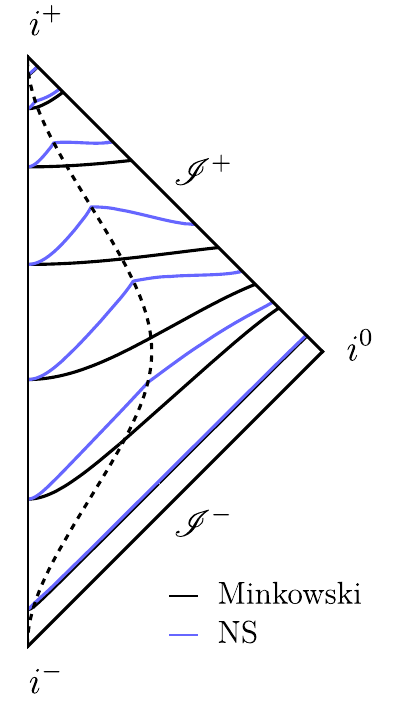}
\vspace{-2ex}
\caption{Conformal Carter-Penrose diagrams depicting foliations of Minkowski (black) and polytropic-like NS (blue) spacetimes, Cauchy on the left and hyperboloidal on the right. The dashed line represents the surface of the star. The hyperboloidal slices of the NS look almost as steep as null slices inside of its surface because the chosen value of $\Kc=-3$ has a large absolute value. Details on the construction of these diagrams are given in \aref{diagrams}.}\label{penns}
\end{figure}

In the interior of the NS, the integral in $int(\rtilde)$ is determined numerically and $\Cc$ is set to zero. Outside of the star's surface, the metric is that of a Schwarzschild BH, so that $e^{\nu(\rtilde)}=e^{-\lambda(\rtilde)}=1-\frac{2\mns}{\rtilde}$ and the boost reduces to the closed-form expression 
\begin{equation}\label{boostexprout}
h'(\rtilde) = -\frac{\left(\frac{\Kc\,\rtilde}{3}+\frac{\Cc}{\rtilde^2}\right)}{\left(1-\frac{2\mns}{\rtilde}\right)\sqrt{\left(1-\frac{2\mns}{\rtilde}\right)+\left(\frac{\Kc\,\rtilde}{3}+\frac{\Cc}{\rtilde^2}\right)^2}} .
\end{equation}
The integration constant is set to $\Cc=0.032978$ (for the present choice of parameters) to ensure continuity of the boost at the star's surface for the chosen parameter values. The resulting boost function is shown on the top plot in~\fref{boost}. 

Hyperboloidal slices are spacelike everywhere in the spacetime, and it is only in the asymptotic limit that they become tangent to null rays. This is reflected in the behaviour of the quantity $-\tilde g_{\tilde t\tilde t}h'(\tilde r)$, which satisfies $|\tilde g_{\tilde t\tilde t}h'(\tilde r)|\leq 1$, as is depicted in the bottom plot in~\fref{boost}.

\Fref{penns} presents the polytropic-like NS data in the form of conformal diagrams, both for Cauchy data determined in~\ssref{cauchytov} from the TOV equations, and for the hyperboloidal data obtained in this subsection. Equivalent slices of the Minkowski spacetime are included for comparison. In the Cauchy slices on the left, the Minkowski and NS cases look indistinguishable, but the difference between both is much clearer on the hyperboloidal slices on the right. Note how the latter intersect \scrip. 

\subsection{Compactification of single neutron star}\label{nscompact}

\begin{figure}[h]
\includegraphics[width=0.95\linewidth]{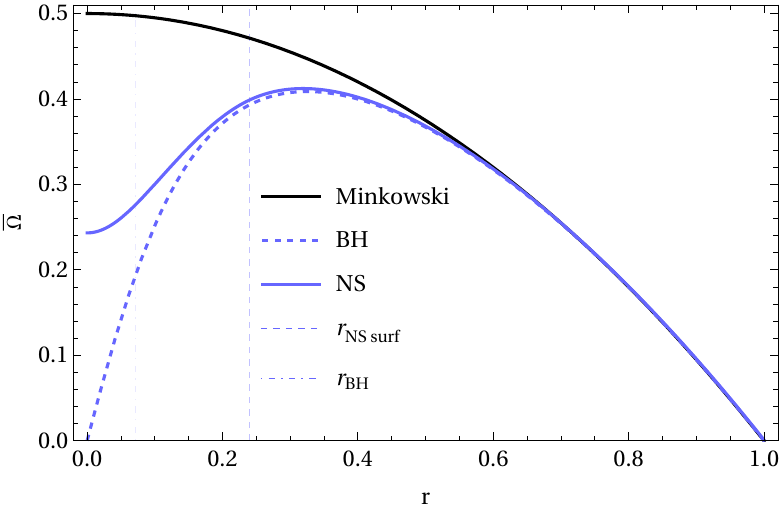}
\vspace{-2ex}
\caption{Compactification factors for Minkowski spacetime ($\aconf=|\Kc|(1-r^2)/6$), for a Schwarzschild BH with mass $\mns$ and for the NS solution. The dashed line represents the surface of the star and the dot-dashed one, the location of the horizon of the BH with the same mass.}\label{compactns}
\end{figure}

To include \scrip\xspace in a finite domain, the infinite-ranged radius $\rtilde$ is compressed into the finite $r$ via
\begin{equation}
\rtilde = \frac{r}{\aconf(r)},
\end{equation}
where $\aconf$ is called the compactification factor. After this coordinate change, line element \eref{metricnshyp} becomes
\begin{eqnarray}\label{metricnshypcomp}
d\tilde s^2 = &&-e^{\nu(r/\aconf)}dt^2 -2e^{\nu(r/\aconf)}h'(r/\aconf)\left(\frac{\aconf-r\aconf'}{\aconf^2}\right)dt\,dr \nonumber \\ && +\left(e^{\lambda(r/\aconf)}-e^{\nu(r/\aconf)}\left[h'(r/\aconf)\right]^2\right)\left(\frac{\aconf-r\aconf'}{\aconf^2}\right)^2dr^2 \nonumber \\ &&+\frac{r^2}{\aconf^2}d\sigma^2 .
\end{eqnarray}
Note that the $\rtilde$ that most functions (except for $\aconf$) depend on has been substituted by $\frac{r}{\aconf}$.
One way to deal with the metric components now diverging at null infinity is conformal compactification \cite{PhysRevLett.10.66,Friedrich:2003fq,Frauendiener2004}. It consists of rescaling the physical metric by a conformal factor $\Omega$ so that the resulting metric $g_{ab} = \Omega^2\tilde g_{ab}$ is finite everywhere. After the conformal rescaling the line element is
\begin{eqnarray}\label{metricnshypcompconf}
ds^2 = &&\Omega^2d\tilde s^2 \nonumber \\ 
= &&-\Omega^2e^{\nu}dt^2 -2\frac{\Omega^2}{\aconf^2}e^{\nu}h'\left(\aconf-r\aconf'\right)dt\,dr \nonumber \\ && +\frac{\Omega^2}{\aconf^2}\left[\left(e^{\lambda}-e^{\nu}\left[h'\right]^2\right)\left(\frac{\aconf-r\aconf'}{\aconf}\right)^2dr^2 \right. \nonumber \\ &&\left.+r^2d\sigma^2\right] .
\end{eqnarray}
One way to determine the compactification factor is by imposing conformal flatness on the spatial part of the metric, which translates to 
\begin{equation}\label{confflatnessns}
\left[\left(1-\frac{2m(\frac{r}{\aconf})\aconf}{r}\right)^{-1}-e^{\nu(\frac{r}{\aconf})}\left(h'(\frac{r}{\aconf})\right)^2\right]\left(\frac{\aconf-r\aconf'}{\aconf}\right)^2=1 .
\end{equation}
This equation is solved for $r\in(0,1)$\footnote{
The value of $\aconf$ at $r=10^{-6}$ is used as guess for the resolution, with the final solution given by that whose $\aconf(1)$ closest to zero. Here $|\aconf(1)|<10^{-7}$ was obtained for $\aconf(10^{-6})=0.243038$.}. The result is shown by the blue solid line in~\fref{compactns}, with the Minkowski and Schwarzschild equivalents for comparison. Note how the value of $\aconf(0)$ of the NS case is between the others, namely a larger value for the empty spacetime and zero for the BH one.  
{The profiles of the evolution variables for the single NS initial data on a compactified hyperboloidal slice are shown in solid blue lines in~\fref{vars}.}

\section{Initial data for a black hole at the center of a neutron star}\label{nsbh} 

The basic steps for putting together a spherically symmetric NS and a small BH at its center and expressing the result on a compactified hyperboloidal slice are: 
\begin{enumerate}\itemsep0em
\item Solve TOV for NS on Cauchy slice (\ssref{cauchytov})
\item Express it in isotropic form (\ssref{nsiso})
\item Add BH in isotropic form (\ssref{addiso})
\item Solve Hamiltonian constraint (\ssref{hsolve}) 
\item Hyperboloidalize (\ssref{bothhyp}) 
\item Compactify (\ssref{bothcompact})
\end{enumerate}

\subsection{Neutron star in isotropic coordinates}\label{nsiso}

This step aims to express the ansatz on Cauchy slices \eref{metricuncomp} solved for in \ssref{cauchytov} in terms of isotropic spatial coordinates as
\begin{equation}
d\tilde s^2 = -e^{\nu}d\ttilde^2+\frac{1}{\chi_\sns}\left(d\rbar^2+\rbar^2d\sigma^2\right) . 
\end{equation}
This requires satisfying the following two relations: 
\begin{subequations}\label{relsisons}
\begin{eqnarray}
e^{\lambda}d\rtilde^2 &=& \frac{d\rbar^2}{\chi_\sns} ,\\
\rtilde^2 &=& \frac{\rbar^2}{\chi_\sns} . 
\end{eqnarray}
\end{subequations}
Eliminating $\chi_\sns$ and reordering gives 
\begin{equation}\label{drtildedrbar}
\frac{d\rtilde}{d\rbar}\equiv\rtilde'(\rbar) = \frac{\rtilde(\rbar)}{\rbar}\frac{1}{\sqrt{e^{\lambda(\rtilde(\rbar))}}} . 
\end{equation}

For the outer spacetime, as the metric is just that of a Schwarzschild BH with mass $\mns$, the conformal factor is just (see~\eref{chibh} for the BH equivalent)
\begin{equation}\label{chinsout}
\chi_{\sns, out} = \left(1+\frac{\mns}{2\rbar}\right)^{-4} . 
\end{equation}
The location of the NS's surface along the isotropic radial coordinate $\rbsurf$ is determined from
\begin{equation}\label{rbsurfdef}
\chi_{\sns, out}(\rbsurf)= \frac{\rbsurf^2}{\rtsurf^2} , 
\end{equation}
where \eref{chinsout} above is to be substituted and $\rtsurf$ was determined when solving the TOV equations in \sref{cauchytov}. 

Equation \eref{drtildedrbar} is solved with the matching boundary condition $\rtilde(\rbsurf)=\rtsurf$.
The solution is used to construct the full NS conformal factor as a piecewise function: \eref{chinsout} for $\rbar>\rbsurf$, and 
\begin{equation}
\chi_{\sns, in} = \frac{\rbar^2}{\rtilde^2} ,
\end{equation}
for the interior of the star $\rbar\leq\rbsurf$ substituting $\rtilde(\rbar)$ with the solution to \eref{drtildedrbar}.
Finally set 
\begin{equation}
\psi_{\sns}\equiv \chi_{\sns}^{-1/4} . 
\end{equation}
This procedure is equivalent to directly solving the TOV equations in isotropic coordinates, as described in Appendix B in~\cite{Tsokaros:2015fea}. 

\subsection{Superposition of neutron star and black hole in isotropic coordinates}\label{addiso}

\noindent The Schwarzschild metric in Schwarzschild coordinates,
\begin{equation}\label{metricschw}
d\tilde s^2 = -\left(1-\frac{2\,\mbh}{\rtilde}\right)d\ttilde^2+\frac{d\rtilde^2}{\left(1-\frac{2\,\mbh}{\rtilde}\right)}+\rtilde^2d\sigma^2, 
\end{equation}
takes the following form in isotropic ones, 
\begin{eqnarray}\label{metricschwiso}
d\tilde s^2 &=& -\left(1-\frac{2\,\mbh}{\rtilde}\right)d\ttilde^2+\frac{1}{\chi_\sbh}\left(d\rbar^2+\rbar^2d\sigma^2\right), \\
&=& -\left(\frac{1-\frac{\mbh}{2\,\rbar}}{1+\frac{\mbh}{2\,\rbar}}\right)^2d\ttilde^2+\left(1+\frac{\mbh}{2\,\rbar}\right)^4\left(d\rbar^2+\rbar^2d\sigma^2\right). \nonumber
\end{eqnarray}
This is obtained by solving equations equivalent to \eref{relsisons} analytically, imposing that $\rbar'(\rtilde=\infty)=1$. The solution corresponds to 
\begin{equation}\label{chibh}
\rtilde = \left(1+\frac{\mbh}{2\,\rbar}\right)^2\rbar \quad \textrm{and} \quad \chi_\sbh = \left(1+\frac{\mbh}{2\rbar}\right)^{-4} . 
\end{equation}
For similarity with the notation used in \cite{Richards:2021upu}, the BH part of the conformal factor will be denoted as
\begin{equation}
\psi_\sbh = \frac{\mbh}{2\rbar} . 
\end{equation}

\begin{figure}[h]
\includegraphics[width=\linewidth]{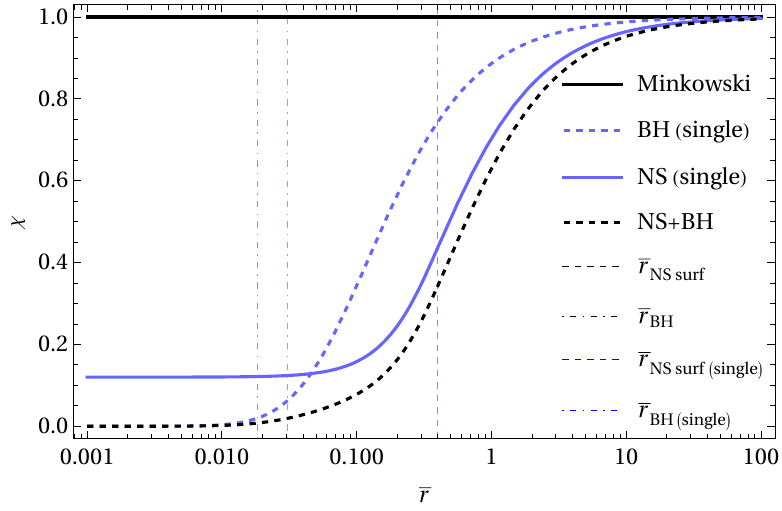}
\vspace{-4ex}
\caption{$\chi$ conformal factor for the single BH $\chi_\sbh$ \eref{chibh}, for the single NS $\chi_\sns$ and for the superposition of both $\chi_{\textrm{\tiny NS+BH}}$ in terms of the isotropic radius. The horizontal axis is in logarithmic scale to better show the lines near the origin.}\label{chisuperp} 
\end{figure}

\begin{figure}[h]
\includegraphics[width=\linewidth]{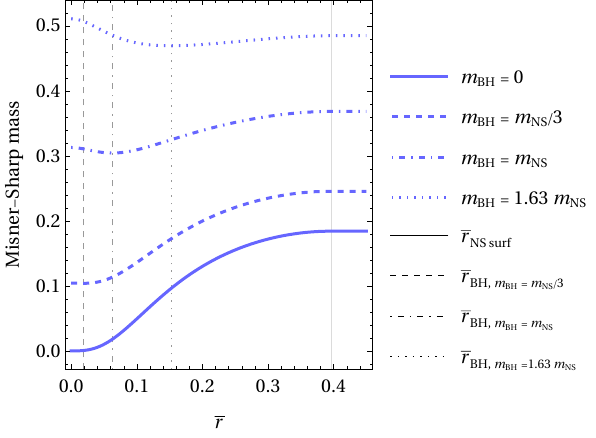}

\includegraphics[width=\linewidth]{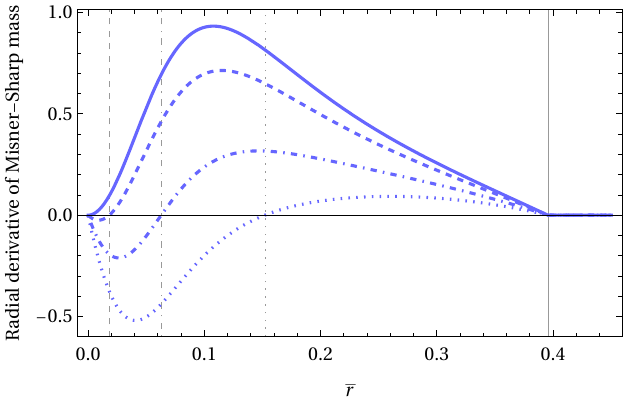}
\vspace{-6ex}
\caption{Misner-Sharp mass (top) and its radial derivative (bottom) for the constraint-violating superposition of the NS with BHs with different masses, specified in the legend that applies to both plots. The solid blue line in the top plot coincides with the black one in~\fref{tovsol}. The effect of the correction that satisfies the Hamiltonian constraint is shown for the $\mbh=\mns/3$ in~\fref{msmassplotsboth}.}\label{msmassplots} 
\end{figure}

The presence of the NS and BH are put together by adding their $\psi$ conformal factors, in a similar way to how Brill-Lindquist initial data \cite{1963PhRv..131..471B} are constructed: 
\begin{equation} \label{psiconstr}
\psi=\psi_\sns + \psi_\sbh + \delta\psi
\end{equation}
The extra term $\delta\psi$ denotes the correction solving the Hamiltonian constraint (see next section), and $\psi$ is such that the spacetime will be asymptotically flat ($\psi(\rbar\to\infty)=1$).  
The first two terms on the right of \eref{psiconstr}, expressed as $\chi_{\textrm{\tiny NS+BH}}=\left(\psi_\sns + \psi_\sbh\right)^{-4}$ are exemplified by~\fref{chisuperp}. 
In the main construction, the BH's mass is set to $\mbh=\mns/3$, which is considerably larger than the BH masses considered in~\cite{Richards:2021upu}. The motivation for this choice is that a large BH mass will more effectively demonstrate robustness of the hyperboloidal construction, as no approximations can be made. It does not make sense to consider a BH whose Schwarzschild radius is equal or larger to the NS's surface (with limiting case shown in blue dotted line in~\fref{msmassplots}), as the NS would be completely enclosed within the BH's horizon. The Misner-Sharp mass function~\cite{Misner:1964je} is a quasi-local measurement of the mass in spherical symmetry, and calculating it for the superposition can illustrate interesting aspects of the spacetime (even if the data are constraint-violating). \Fref{msmassplots} shows the features of the Misner-Sharp mass\footnote{Calculated following subsection~2.6 in~\cite{Vano-Vinuales:2015lhj} with $\Omega=1$.} for several configurations of the superposition. Very small BH masses would be very close to the blue solid line (only NS) and the location of their horizons would be indistinguishably close to the origin. For larger values of the BH mass the horizons move further towards larger radii as expected, and they are located at the minimum of their corresponding Misner-Sharp mass, which increases inside of the horizon. This is more clearly seen in the bottom plot: the radial derivative of the Misner-Sharp mass is negative inside of the horizon for the superpositions. The effect is small for $\mbh=\mns/3$, but it is clearly more pronounced for larger BH masses. Having a Misner-Sharp mass that is not monotonically increasing from the origin outwards, but which actually increases inside of the horizon, causes problems with the recovery of the solution in terms of the areal radius (see~\ssref{ssareal}) within the horizon.

\subsection{Resolution of Hamiltonian constraint}\label{hsolve}

The resolution of the Hamiltonian constraint follows the procedure in~\cite{Richards:2021upu} (where $\delta\psi$ is denoted as $u$), of which the main steps are included. The superposed initial data constructed in the previous subsection are such that the momentum constraint is already satisfied. The Hamiltonian constraint in terms of $\psi=\chi^{-1/4}$ takes the form
\begin{equation} \label{Hbasic}
\Delta \psi = -2\pi\psi^5\rho,
\end{equation}
with $\Delta$ the spatial Laplacian and $\rho=\tilde n^a\tilde n^bT_{ab}$ the density as measured by a normal observer (ADM density), and in this setup $\rho\equiv\enerdens$ as the pressure part cancels. Using the conformal freedom in rescaling the density, substitute 
\begin{equation} \label{rhoresc}
\rho=\psi^m\bar \rho 
\end{equation}
with the choice $m=-6$, which will later eliminate the divergence at the origin that appears due to the BHs presence. After substituting \eref{psiconstr}, 
\begin{equation}
\Delta\psi_\sns=-2\pi\psi_\sns^5\enerdens_\sns , \qquad \Delta\psi_\sbh=0, 
\end{equation} 
and 
\begin{equation}
\bar \rho = \psi_\sns^{-m}\enerdens_\sns 
\end{equation}
into \eref{Hbasic} with \eref{rhoresc}, we arrive at the differential equation to solve for $\delta\psi$: 
\begin{eqnarray}\label{Hfinal}
\Delta\delta\psi&=&2\pi\enerdens_\sns\left(\psi_\sns^5-\frac{\left(\psi_\sns + \psi_\sbh + \delta\psi\right)^{5+m}}{\psi_\sns^m}\right), \nonumber \\
\partial^2_\rbar\delta\psi+\frac{2}{\rbar}\partial_\rbar\delta\psi &=& 2\pi\enerdens_\sns\left(\psi_\sns^5-\frac{\psi_\sns^6}{\left(\psi_\sns + \psi_\sbh + \delta\psi\right)}\right),
\end{eqnarray}
where in the second line spherical symmetry has been assumed and $m=-6$ has been set. 

\begin{figure}[h]
\includegraphics[width=0.95\linewidth]{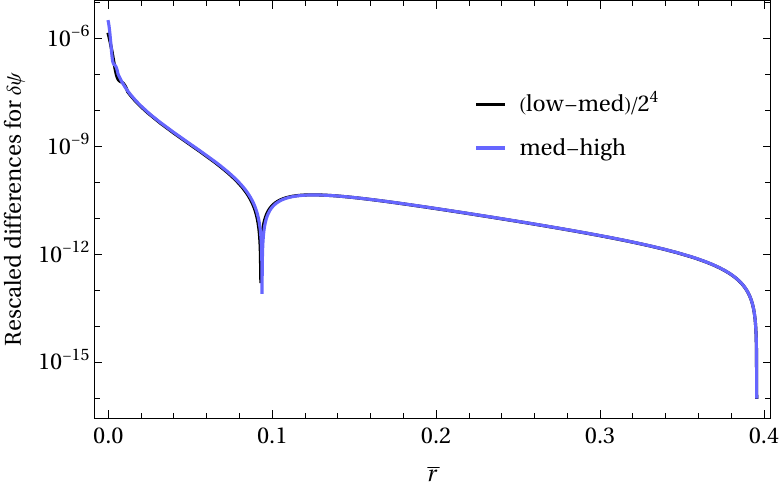}
\vspace{-3ex}
\caption{Pointwise convergence plot showing the rescaled differences for $\delta\psi$, with ``low'', ``med'' and ``high'' respectively denoting $\delta\psi$ with 100, 200 and 400 spatial points. The solutions were integrated with a 4th order Runge-Kutta starting at the right boundary using the value of $A$ found by tuning in the \texttt{NDSolve} results, and the solutions were interpolated using 6th order. The mass of the BH is $\mbh=\mns/3$.}\label{conv}
\end{figure}

Outside of the NS's surface, $\delta\psi$ must take the functional form $A/\rbar$, with $A$ a real constant, to satisfy asymptotic flatness and the homogeneous Laplace equation. This provides the conditions $\delta\psi(\rbsurf)=A/\rbsurf$ and $\partial_\rbar\delta\psi(\rbsurf)=-A/\rbsurf^2$ to impose when solving \eref{Hfinal}. Tuning the value of the constant $A$ allows the integrator to find a solution for $\delta\psi$ that is even at the origin. The differential equation \eref{Hfinal} is solved using this condition 
with a 4th-order Runge-Kutta integrator. This was done to obtain an estimate of the convergence of the solution, shown in~\fref{conv}. Indeed the expected 4th order convergence is recovered\footnote{Note however that the solution is constructed on \texttt{InterpolatedFunction}s of the NS energy density and mass profiles obtained  with \texttt{Mathematica}'s \texttt{NDSolve}, so there are probably sources of error that are not being taken into account. Still, this convergence serves as an indication that the solution is reliable.}.

\begin{figure}[h]
\includegraphics[width=0.95\linewidth]{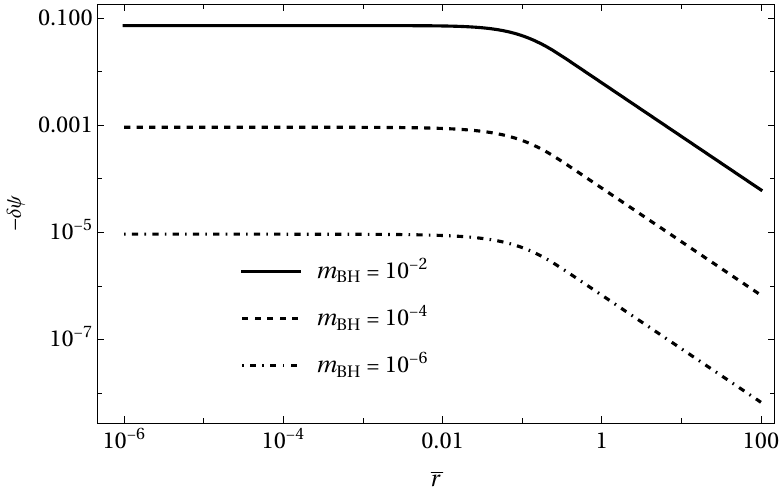}
\vspace{-3ex}
\caption{Minus $\delta\psi$ in double logarithmic scale as a function of the isotropic radial coordinate for several values of the central BH's mass. The tuned values of $A$ for the displayed solutions are, from larger to smaller BH masses, $-6.6225\cdot 10^{-3}$, $-6.7379\cdot 10^{-5}$ and $-6.7443\cdot 10^{-7}$. Plotting the solutions in linear scale would provide something qualitatively similar to the dashed green line in figure~9 in~\cite{Richards:2021upu}.}\label{deltapsisol} 
\end{figure}

\begin{figure}[h]
\includegraphics[width=0.95\linewidth]{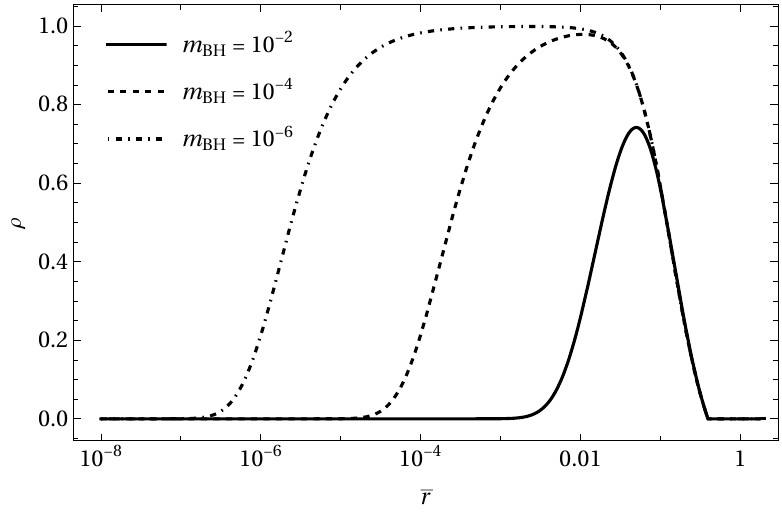}
\vspace{-2ex}
\caption{ADM density~\eref{rhoresc} for different masses of the central BH as a function of the isotropic radius in logarithmic scale using $\polyK=1$ and $\Gamma=2$. This behaviour is qualitatively the same as that for the rest-mass density in~\cite{Richards:2021upu}'s figure~2.}\label{rhosol}
\end{figure}

Figure~2 in \cite{Richards:2021upu} shows the effect of varying the mass of the BH (for a fixed value of the NS's mass) on the rest-mass density. Here three different values\footnote{Smaller than $\mbh=\mns/3\approx 0.06$ to better compare to~\cite{Richards:2021upu}.} of $m_\sbh=10^{-2},10^{-4},10^{-6}$ have been considered, and their (negative) solutions $\delta\psi$ are displayed with opposite sign and logarithmic scale in~\fref{deltapsisol}. Their corresponding densities~\eref{rhoresc} are shown in~\fref{rhosol}. 

\begin{figure}[h]
\includegraphics[width=0.555\linewidth]{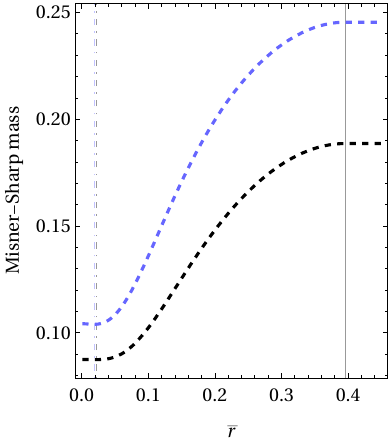}
\includegraphics[width=0.425\linewidth]{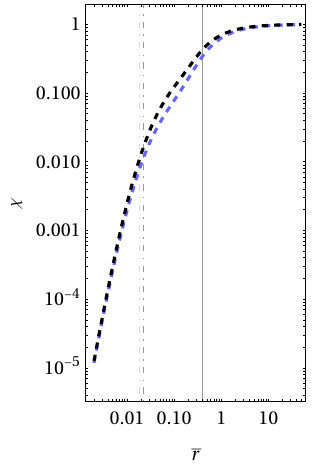}

\includegraphics[width=\linewidth]{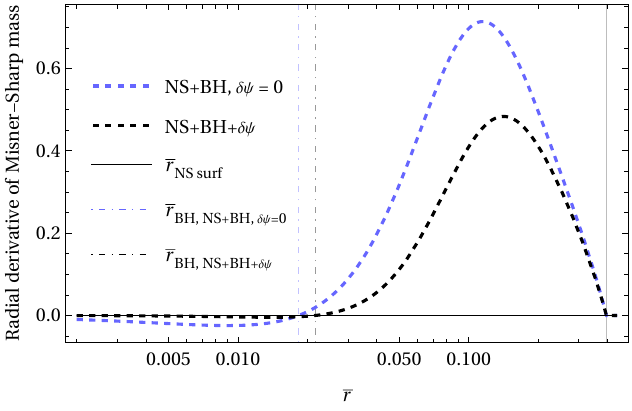}
\vspace{-6ex}
\caption{Properties of the superposition with $\mbh=\mns/3$ (and $\polyK=1$ and $\Gamma=2$ for the NS part) with and without correction $\delta\psi$ solving the Hamiltonian constraint, respectively in black and blue. The top-right plot shows the difference in the spatial conformal factor $\chi$ for both cases. The top-left plot shows the Misner-Sharp mass, and the bottom one, its radial derivative. The legend applies to all three plots.}\label{msmassplotsboth} 
\end{figure}

The solution to the Hamiltonian constraint setting $\mbh=\mns/3$ is the one that will be used for the hyperboloidalization and compactification in the rest of the paper. The profile of its Misner-Sharp mass is shown in dashed black in the top-left plot in~\fref{msmassplotsboth}: indeed the values for the corrected superposition are smaller than the constraint-violating ones (in dashed blue) because the $\delta\psi$ correction is negative. The latter can be interpreted as the binding energy of the system, such that the total energy is smaller for the constraint-satisfying solution. The horizon is denoted by a vertical black dot-dashed line for the case with $\delta\psi$ correction, and it is located at a larger value of the radius than for the constraint-violating superposition (horizon in dot-dashed blue). See for this the bottom plot, where the radial derivative of the Misner-Sharp mass shows a less pronounced increase of the mass inside the horizon for the constraint-satisfying case. The spatial conformal factor $\chi=\psi^{-4}$~\eref{psiconstr}, shown on the top-right, does not show a significant difference between the two superpositions.

\vspace{-4ex}

\subsection{Hyperboloidalization of superposition}\label{bothhyp}

The present procedure of expressing the metric of interest on a hyperboloidal slice has as goal to be used with conformal compactification in numerical evolutions. Conformal compactification relies on the asymptotic interplay between the spatial compactification $\aconf$ and the conformal rescaling of the metric with $\Omega$ (see \eref{metricnshypcomp} and \eref{metricnshypcompconf} respectively for the NS case), so that close to \scrip, $\Omega\sim\aconf$. This is more straightforward if the metric being hyperboloidalized is expressed in terms of the areal radius, so that 
\begin{equation}\label{confresclinenemboth}
d\tilde s^2 = [...] +\rtilde^2d\sigma^2 = [...] +\frac{r^2}{\aconf^2}d\sigma^2 \ \to \ ds^2=[...] + \frac{\Omega^2}{\aconf^2}r^2d\sigma^2 . 
\end{equation}
Thus, the first step when hyperboloidalizing will be to transform the NS+BH superposed metric satisfying the Hamiltonian constraint from being expressed in terms of the isotropic radius $\rbar$ to the areal one $\rtilde$. 

\subsubsection{Transformation to areal radius}\label{ssareal}

Let us focus on the spatial part of the line element,
\begin{equation}
d \tilde l^2 = \psi^4\left(d\rbar^2+\rbar^2d\sigma^2\right) = \B\,d\rtilde^2+\rtilde^2d\sigma^2 , 
\end{equation}
where it is first expressed in terms of the isotropic radius and after the second equality in terms of the areal one. In the opposite way as done in \ssref{nsiso}, the aim is to determine the expression for $\B$ to express the metric in terms of $\rtilde$. Equating each term in the line element yields (assuming that $\B$ is positive)
\begin{subequations}\label{relsisoareal}
\begin{eqnarray}
\psi^2d\rbar &=& \sqrt{\B}d\rtilde, \label{relsisoareal1}\\
\psi^2\rbar&=&\rtilde . \label{relsisoareal2}
\end{eqnarray}
\end{subequations}
Deriving~\eref{relsisoareal2} and substituting it in~\eref{relsisoareal1} gives for~$\B$
\begin{equation}\label{Beq}
\B = \left(\frac{\psi}{2\,\rbar\,\partial_\rbar\psi+\psi}\right)^2 . 
\end{equation}
Now $\B$ has to be expressed in terms of $\rtilde$. Outside of the NS's surface, the total conformal factor is 
\begin{equation}
\psi = 1+\frac{m_\sns}{2\rbar}+\frac{m_\sbh}{2\rbar}+\frac{A}{\rbar}\equiv 1+\frac{M}{2\rbar} ,
\end{equation}
with 
\begin{equation}\label{totalmass}
M=m_\sns+m_\sbh+2A  
\end{equation}
the total mass contained in the system and coinciding with the Misner-Sharp mass outside of the NS.
Substituting this into~\eref{relsisoareal2} yields the close-form expression for $\rtilde$ (for the solution where $\rbar\sim\rtilde$ asymptotically)
\begin{equation}
\rbar = \frac{1}{2} \left(\sqrt{\rtilde (\rtilde-2M)}-M+\rtilde\right). 
\end{equation}
This yields $\B=1/\left(1-\frac{2M}{\rtilde}\right)$, as expected for the spacetime outside of the NS. 
Between the horizon of the central BH $\rbhor$ and the NS surface, $\rbar(\rtilde)$ is calculated numerically. The procedure involves interpolating $\left\{\rtilde,\bar{r}\right\}=\left\{(\psi_\sns(\rbar)+\psi_\sbh(\rbar)+\delta\psi(\rbar))^2\rbar,\rbar\right\}$ for $\rbar\in[\rbhor,\rbsurf]$, where $\rbhor$ is the solution to $1/\B(\rbar)=0$ setting~\eref{Beq}, and determining $\rbsurf$ from~\eref{rbsurfdef}. The difficulty lies in finding $\rtilde(\rbar)$ inside of the BH's horizon $\rbhor$. This is because inside of the horizon $\B<0$ and the isotropic radial coordinate $\rbar$ only covers the outside spacetime -- see \aref{isoareal} for more details. However, to obtain a hyperboloidal trumpet slice of the NS+BH superposed spacetime, $\rtilde$ needs to reach within the horizon. 

\begin{figure}[h]
\includegraphics[width=0.95\linewidth]{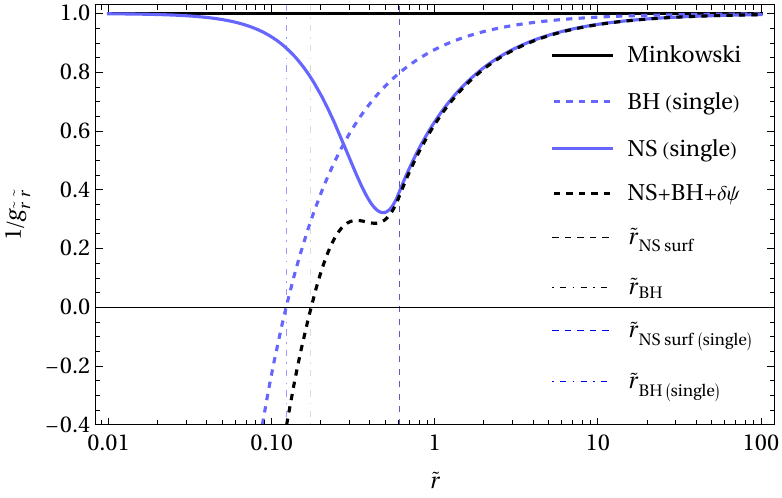}
\vspace{-2ex}
\caption{$\B^{-1}$ in terms of the areal radius $\rtilde$ after ``undoing'' the change to isotropic coordinates. The result of this subsection's calculations is the thick black dashed line, the others are included for comparison. The $x$ axis is set in logarithmic scale to better show the features of the construction.}\label{oogrrareal}
\end{figure}

Full details of how $\rbar(\rtilde)$ and $\B(\rbar(\rtilde))$ are calculated inside of the horizon, illustrated with a spacetime including only a BH, are given in \aref{isoareal}. The summary of the steps followed\footnote{This procedure works for as long as the Misner-Sharp mass of the system is monotonically increasing or very close to it.} is:
\begin{enumerate}\itemsep0ex
\item Interpolate $\left\{\rtilde,\bar{\bar r}\right\}=\left\{\rthor\frac{1-\cos(\pi x)}{2},\rbhor\sin(\pi x)\right\}$ between~$x\in[0,1]$, with $\rthor=(\psi_\sns(\rbhor)+\psi_\sbh(\rbhor)+\delta\psi(\rbhor))^2\rbhor$ corresponding to the vertical black dot-dashed line in~\fref{oogrrareal}. This quantity is denoted as $\bar{\bar r}(\rtilde)$ and is the imaginary part of the isotropic radius in terms of the areal one, shown as a black dashed line in~\fref{isoellipse} for Schwarzschild. 
\item Construct the isotropic radial coordinate as
\begin{equation}\label{complexrbar}
\rbar=i\,\bar{\bar r}(\rtilde) +\rbhor\left(\frac{\rtilde}{\msm(\rtilde=10^{-6})}-1\right) , 
\end{equation}
with $\msm$ the Misner-Sharp mass, evaluated virtually at the origin. This evaluates to complex values in the region of interest $\rtilde\in(0,\rthor)$. 
\item Evaluate $\psi(\rbar(\rtilde))$ as~\eref{psiconstr} inside of the horizon $\rbar\in(0,\rbhor)$ using~\eref{complexrbar} above\footnote{\texttt{InterpolatedFunction}s in \texttt{Mathematica} that interpolate real values (the format in which most numerical solutions were given) cannot be evaluated on complex numbers. Thus, expressing $\chi_\sns=\psi_\sns^{-4}$ and $\delta\psi$ as \texttt{InterpolatingPolynomial}s (with 20 points and for $\rbar\in(0,\rbhor)$) was needed for the substitution of~\eref{complexrbar} into $\psi(\rbar(\rtilde))$ given by~\eref{psiconstr}.}. The final profile for $\B(\rtilde)$ inside of the horizon is the real part of~\eref{Beq}. 
\end{enumerate}
The curve describing $\B^{-1}$ in terms of the areal radius $\rtilde$ is shown by the black dashed line in~\fref{oogrrareal}. 

\subsubsection{Construction of boost function}

So far in this section we put together a NS and a BH in isotropic coordinates, obtained the correction $\delta\psi$ that solved the Hamiltonian constraint, and expressed the result in terms of the areal radius. This has comprised the spatial part of the metric, but nothing has been stated yet regarding its $\tilde g_{\ttilde\ttilde}$ component. 
If considering quasi-equilibrium initial data, it would make sense to construct $\tilde g_{\ttilde\ttilde}$ by solving the extended conformal thin-sandwich equations~\cite{Pfeiffer:2002iy,York:1998hy}. However, a parasitic BH about to accrete parts of a NS is hardly a configuration in equilibrium. As the choice of gauge (interpreted here as the choice of $\tilde g_{\ttilde\ttilde}$) does not contain physical degrees of freedom, there is a certain liberty here. Possibly the simplest and most convenient option is
\begin{equation}\label{gttchoice}
\tilde g_{\ttilde\ttilde}\doteq -\frac{1}{\tilde g_{\rtilde\rtilde}} \equiv -\Aa(\rtilde) ,
\end{equation}
in similarity with the metric structure of a spherically symmetric BH e.g.~\eref{metricschw}. 

Condition~\eref{cmccond}, to be satisfied for CMC hyperboloidal slices, is another reason for choosing the above expression for $\tilde g_{\ttilde\ttilde}$, because then the determinant of the metric is simply $\tilde g=-\rtilde^4\sin^2\theta$, as compared to~\eref{gnns}. The quantity $int(\rtilde)$ defined in ~\eref{hkint} integrates analytically to $-\frac{\Kc \rtilde}{3}-\frac{\Cc}{\rtilde^3}$ and~\eref{boostexprin} becomes
\begin{equation}\label{boostexprboth}
h'(\rtilde) = -\frac{\left(\frac{\Kc \rtilde}{3}+\frac{\Cc}{\rtilde^3}\right)}{\Aa(\rtilde)\sqrt{\Aa(\rtilde)+\left(\frac{\Kc \rtilde}{3}+\frac{\Cc}{\rtilde^3}\right)^2}} ,
\end{equation}
where $\Aa(\rtilde)$ is substituted by $1/\B(\rtilde)$ as obtained above. 

\begin{figure}[h]
\includegraphics[width=0.95\linewidth]{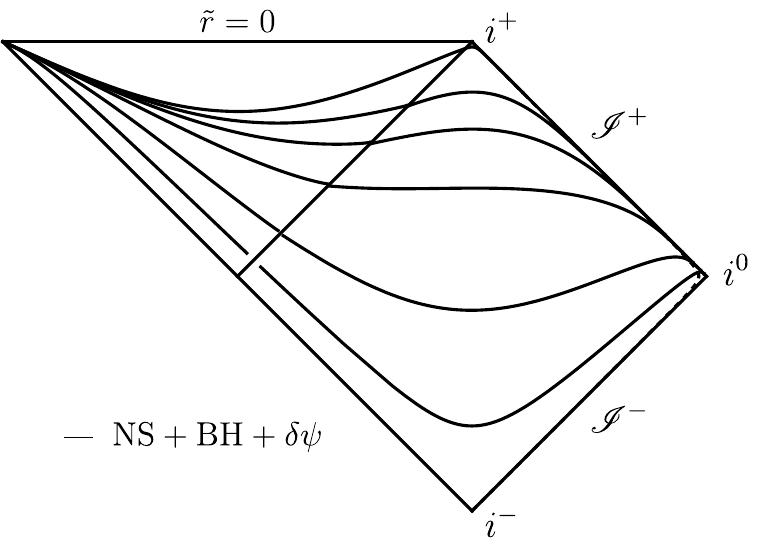}
\vspace{-2ex}
\caption{Conformal diagram presenting the CMC hyperboloidal slices determined for the NS+BH plus correction $\delta\psi$. Details on the construction of the diagram are given in \aref{diagrams}. The NS surface is denoted by a black dashed line that can barely be see at the right of the diagram, close to $i^0$; it is the compactification along the null directions in~\eref{basic2} and the choice of $B=4\,m_{\textrm{\tiny BH superposition}}$ in~\eref{ed:TR} that positioned it almost at the visible edge of the diagram. Modifying the compactification to make it more visible did not add to the information shown. The small discontinuity in the lowest slice is due to the imperfect integration of the height function near the horizon. Different ways to improve it are given in \cite{Vano-Vinuales:2023pum}.}\label{penboth}
\end{figure}

The remaining task is to find the value of $\Cc$ that provides a trumpet slice, see discussion in~\cite{Vano-Vinuales:2023yzs}. The appropriate value is found by analysing the boost function. The above construction is such that along slices of a spacetime with a BH, the quantity $\Aa(\rtilde)h'(\rtilde)$ can i) be finite and continuous, in which case the slices will intersect the singularity; ii) take complex values in a region of the domain, where the slices are not defined; and iii) be real everywhere and have a divergence at some value of the radial coordinate, which corresponds to the location of the trumpet $\rtilde\trum$. Thus, to obtain the $\Cc$ that provides a trumpet, the value of $\Cc$ was tuned to obtain a divergence in $\Aa(\rtilde)h'(\rtilde)$ for a single value of the radius\footnote{For $\polyK=1$, $\Gamma=2$, $\enerdens_{\textrm{center}}=1$ (the present NS configuration), $\mbh=\mns/3$ and $\Kc=-3$, the value $\Cc=0.012345$ makes $\Aa(\rtilde)h'(\rtilde)$ diverge beyond $-8000$ at~$\rtilde\trum=0.14483$.}. 
The slices corresponding to the constructed boost are shown in~\fref{penboth}. 

\subsection{Compactification of superposition}\label{bothcompact}

Compactification of the hyperboloidal slices constructed above can be performed similarly as in~\ssref{nscompact}. For that, introduce $h'(\rtilde)$ \eref{boostexprboth} and $\Aa(\rtilde)$ \eref{gttchoice} in the generic version of~\eref{confflatnessns} to be solved, namely 
\begin{equation}\label{confflatnessboth}
\frac{1-\left[\Aa(\frac{r}{\aconf}) h'(\frac{r}{\aconf})\right]^2}{\Aa(\frac{r}{\aconf})}\left(\frac{\aconf-r\aconf'}{\aconf}\right)^2=1 .
\end{equation}

In this case it is useful to solve for $\aconf$ in two parts: outside and inside of the NS's surface. As the outside spacetime corresponds to a Schwarzschild geometry, set $\Aa(\rtilde)=1-\frac{2M}{\rtilde}$, so that the asymptotic behaviour is that corresponding to the total mass in the system. Setting the chosen value of $\Kc=-3$ and the $\Cc$ found above, with $\aconf_{out}(r=1)=0$ as boundary condition, provides a solution $\aconf_{out}$ that does not cover until $r=0$ but is correct outside of the NS. The location of the NS's surface $\rsurf$ in terms of the compactified coordinate is found solving $\rtsurf =\rsurf/\aconf_{out}(\rsurf)$. The interior solution $\aconf_{in}$ is found with the continuity condition $\aconf_{in}(\rsurf)=\aconf_{out}(\rsurf)$. 

\begin{figure}[h]
\includegraphics[width=0.95\linewidth]{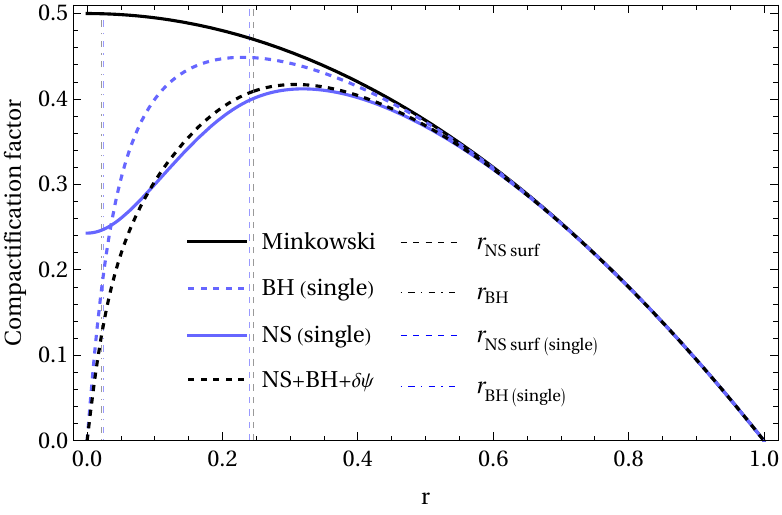}
\\
\includegraphics[width=0.95\linewidth]{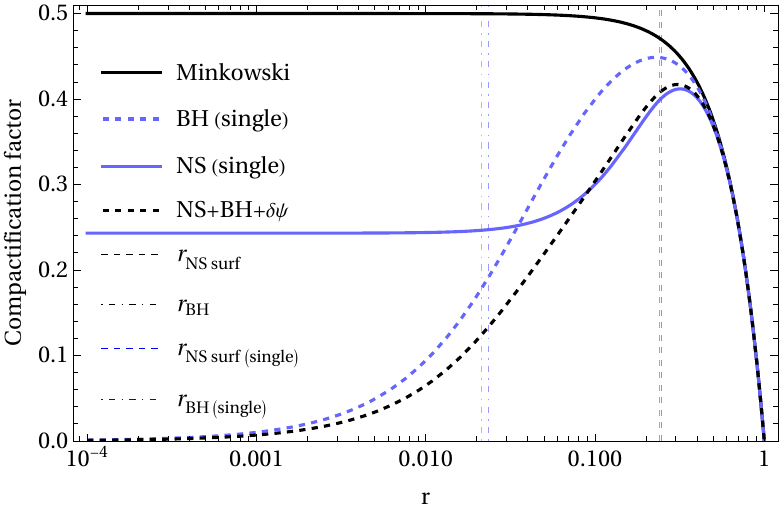}
\vspace{-2ex}
\caption{Compactification factor for the superposition of NS+BH with correction $\delta\psi$, plotted with a black dashed line. 
The other curves, corresponding to $\aconf=|\Kc|(1-r^2)/6$ for Minkowski spacetime, the NS compactification factor found in~\sref{ns} and the same for a single BH with the same mass as the superposed one, are included for ease of comparison and interpretation, same as the location of the horizons and the surfaces of the NSs. As compared to the upper plot, where the horizontal axis in linear in the compactified radial coordinate $r$, the lower one is plotted in logarithmic scale and allows to better see the curves near the origin, which corresponds to the location of the trumpet in the spacetimes including a BH.}\label{aconfboth}
\end{figure}

The final complete result for the compactified hyperboloidal slice is presented in black dashed in~\fref{aconfboth}. The same curves are shown using a linear (above) and log (below) scale in the horizontal axis, to allow to better see the different features of the solution. 
{The black dashed curves in~\fref{vars} show the profiles of the non-trivial initial data for the evolution variables.}

\section{Conclusions}\label{conclu}

The goal of this work was the successful construction of hyperboloidal initial data for spacetimes including a NS, and optionally a BH, in spherical symmetry. Indeed it has been achieved using smooth CMC hyperboloidal slices that were then radially compactified to provide initial data ready for evolution{, as presented in \aref{inidata}}. The described derivations are available as a \texttt{Mathematica} notebook in~\cite{nsbhrepo}. To the author's best knowledge, this is the first time that a NS spacetime is expressed on a hyperboloidal slice. This project has considered a simple polytropic-like equation of state as simplest NS model. However, as the construction is performed numerically, the procedure is extensible to proper polytropes and other numerical equations of state, such as piecewise-polytropic~\cite{Read:2008iy,OBoyle:2020qvf} or tabulated ones, e.g.~\cite{Steiner:2012rk,Gulminelli:2015csa}. 
Only CMC hyperboloidal slices were considered in the construction, but other options can also be tested. As a related example, the hyperboloidal layer construction~\cite{Zenginoglu:2010cq,Bernuzzi:2011aj,Vano-Vinuales:2024tat} (inner Cauchy part matched with an outer hyperboloidal one) will be considered for a NS spacetime in future efforts. 

This work represents a first step towards the generic description of hyperboloidal slices of spacetimes including NSs and/or BHs, of which the most relevant scenario is a binary configuration (such as~\cite{Buchman:2009ew,schinkelthesis} for two BHs). 
The Misner-Sharp mass provides a quasi-local value of the energy contained within a sphere, and it has been especially useful in this project for understanding the superposition of NS and BH, having revealed how the mass' value increased within the horizon and confirmed that, as expected, \eref{totalmass} is indeed the total mass contained in the spacetime. The Misner-Sharp mass is however not defined beyond spherical symmetry, where there does not exist a generic notion of quasi-local mass or energy. Relating to the discussion after~\eref{ttrafo}, to build the correct boost function for the hyperboloidal construction the total mass in the spacetime needs to be known. The configurations of interest correspond to asymptotically flat spacetimes, for which the ADM mass or the Hawking quasi-local mass~\cite{Hawking:1968qt} (which asymptotes to the ADM mass) can be calculated. This should ensure the correct behaviour near \scrip\xspace for the hyperboloidal slice, even if building it in the interior region will probably be more complicated. 
{While no conceptual limitations are apparent to considering setups beyond spherical symmetry, their main practical difficulty is that expressions and equations will depend not only on the radius, but on two or three coordinates, making the construction and resolution procedures far more involved. In the specific and interesting case of the inclusion of a spinning BH, first a hyperboloidal trumpet description of the Kerr spacetime needs to be developed -- possibly building on~\cite{Schinkel:2013tka,Schinkel:2013zm}.}

The next natural step for this project is the addition of the relativistic Euler equations to the Einstein equations in spherical symmetry for evolving the hyperboloidal initial data derived here. In three-dimensional general relativistic hydrodynamic simulations, the vacuum outside of the stars is customarily filled with a very low density ($<=10^{-7}$), which is referred to as the ``atmosphere''. This allows to properly treat the matter fluids. However, for a compactified slice (like those used for hyperboloidal evolutions), even a very scarce atmosphere will amount to an infinite mass contained in the domain, potentially intractable from an implementation perspective. The setup in~\cite{Duez:2008rb} uses a second grid for the hydrodynamical quantities that covers only the region including NS matter. This inspires an option worth pursuing: the gradual deactivation of the atmosphere close to null infinity. Matter cannot travel at the speed of light, so it is unable to reach \scrip\xspace anyway, which makes this a promising approach. Simulations in spherical symmetry may even be simpler than this: as it is relatively straightforward to keep track of the location of the NS's surface, it is enough to only evolve the hydrodynamic equations inside of it, so the atmosphere is not needed. 

{The first setup to evolve on hyperboloidal slices will be the single NS. While not very physically relevant, the long-term evolution of the static equilibrium solution will be an important first test for the implementation's robustness. In case the NS solution used is gravitationally unstable (like the one considered here), its collapse to a BH will be studied on hyperboloidal slices. To the best of the author's knowledge, collapse on hyperboloidal slices has so far only been studied with scalar fields~\cite{Vano-Vinuales:2016mbo,Peterson:2024bxk,Alvares:2025jjm}, so a more realistic matter collapse scenario is very exciting. Perturbations on top of the static NS initial data will allow to test the implementation's ``atmosphere'' setup more stringently and study how well the hyperboloidal approach deals with matter propagating further from the center of the domain. Evolutions will at first be restricted to spherical symmetry, so the richer phenomenology of e.g. axisymmetric pressure perturbations exciting even-parity f-modes~\cite{Baiotti:2008nf} in the NS, or specific gravitational field perturbations giving rise to odd-parity modes~\cite{Pazos:2006kz} that then excite w-modes~\cite{Kokkotas:1992ka} will take some time to become available. To start with, radial pressure perturbations will be considered. Another possibility is to include a massless scalar field radial perturbation to drive the dynamics. It may be possible to consider a massive scalar field in the central part of the domain (it needs to be massless at \scrip), in which case it could be interpreted as dark matter in the NS spacetime. All these setups will require solving at least the Hamiltonian constraint equation for constraint-satisfying initial data. 
Once this is understood, the NS + small BH superposition will be evolved to study Bondi accretion as a hyperboloidal version of~\cite{Richards:2021upu}. There may be chances to either solve for quasiequilibrium sequences for part of the adiabatic Bondi accretion process, or to reach them after a short evolution from initial data. This would allow to compare to and gain further insights into the actual numerical evolution of the whole process.  
}

\acknowledgments 

The author would like to thank Carlos Palenzuela, Hannes R\"uter and Koutarou Kyutoku for valuable and useful comments on the manuscript. 
The author also gratefully thanks Koutarou Kyutoku for the motivation for this work, and for valuable discussions and feedback on its progress. Acknowledged is also the hospitality of the Department of Physics of Kyoto University, where this work was started. 

The visit was supported by the European Union's Horizon 2020 research and innovation programme under the Marie Sklodowska-Curie grant agreement No 101007855. 
The author thanks the Fundac\~ao para a  Ci\^encia e Tecnologia (FCT), Portugal, for the financial support to the Center for Astrophysics and Gravitation (CENTRA/IST/ULisboa) through the Grant Project~No.~UIDB/00099/2020. Funding with DOI 10.54499/DL57/2016/CP1384/CT0090 is graciously acknowledged. 
This work was also supported by the Universitat de les Illes Balears (UIB); the Spanish Agencia Estatal de Investigación grants PID2022-138626NB-I00, RED2022-134204-E, RED2022-134411-T, funded by MICIU/AEI/10.13039/501100011033 and the ERDF/EU; and the Comunitat Autònoma de les Illes Balears through the Conselleria d'Educació i Universitats with funds from the European Union - NextGenerationEU/PRTR-C17.I1 (SINCO2022/6719) and from the European Union - European Regional Development Fund (ERDF) (SINCO2022/18146).

\appendix

\section{Recovery of the areal radius from the isotropic one with a black hole}\label{isoareal}

\begin{figure}[h]
\includegraphics[width=0.97\linewidth]{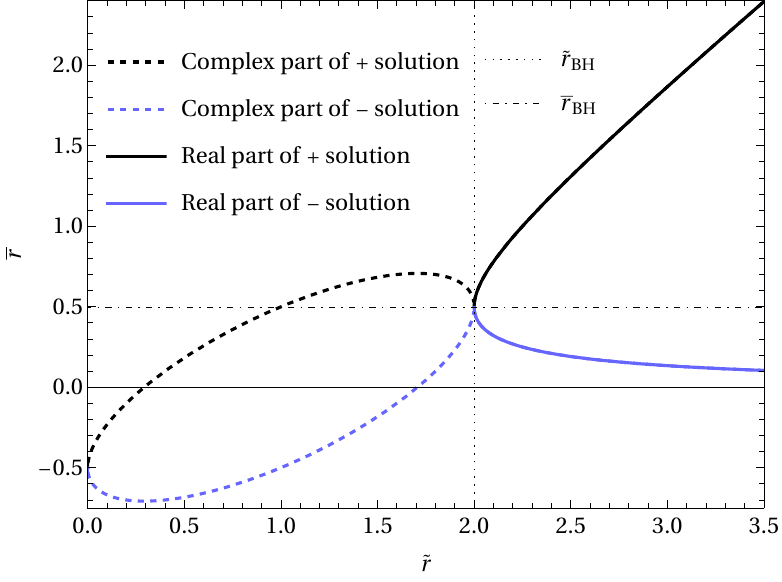}
\vspace{-2ex}
\caption{Isotropic radial coordinate as a function of the areal one for a single BH with unity mass. The thick lines correspond to the solutions~\eref{isosol} with the $+$ (black) and $-$ (blue) lines. The complex part of the relation inside of the horizon is represented with dashed lines. The thin lines locate the positions of the horizon in each of the coordinates.}\label{isoschw}
\end{figure}

The transformation to isotropic coordinates for a Schwarzschild BH as given by~\eref{chibh} is valid outside of the horizon where the~$\rtilde\rtilde$ component of its metric~\eref{metricschw} is positive. The range $\rbar\in(0,\infty)$ covers twice the values of the Schwarzschild coordinate outside of the horizon (for $\rtilde\geq \rthor = 2\mbh$, see solid lines in~\fref{isoschw}), but it does not reach inside of the horizon. This is relevant when transforming from isotropic to Schwarzschild or areal-type spatial coordinates and having $\B^{-1}$ change sign and behave inside of the horizon as expected. 

The areal radial coordinate isolated from~\eref{chibh} yields
\begin{equation}\label{isosol}
\rbar = \frac{1}{2}\left(\rtilde-\mbh\pm\sqrt{\rtilde(\rtilde-2\mbh)}\right). 
\end{equation}
These solutions are indeed real for $\rtilde\geq 2\mbh$, and complex for $\rtilde< 2\mbh$. They are represented in~\fref{isoschw} for $\mbh=1$. 

Calculations with explicit closed-form metric solutions have no difficulty, but doing the equivalent computation for a numerically-determined metric, as is required in~\ssref{bothhyp} for the hyperboloidalization of the constraint-satisfying superposition of NS and BH, is more challenging. The procedure summarized in~\ssref{ssareal} is motivated by the following considerations. 

\begin{figure}[h]
\includegraphics[width=0.97\linewidth]{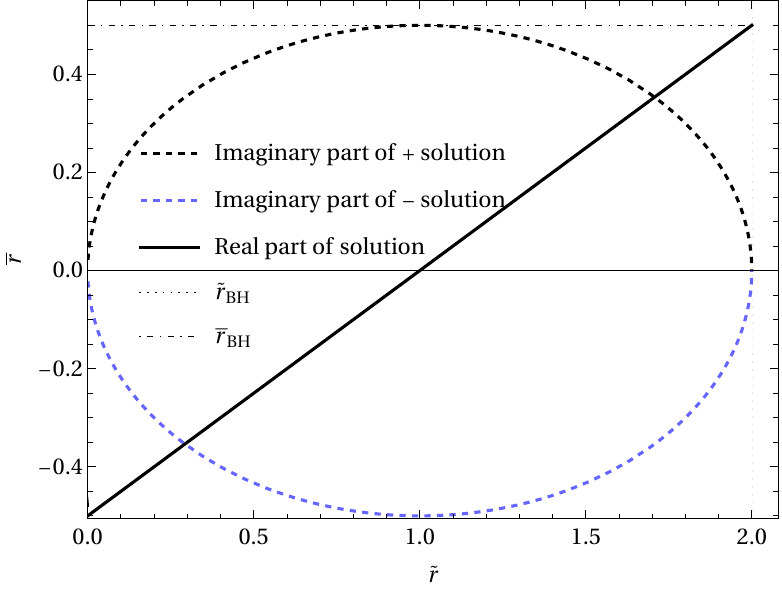}
\vspace{-2ex}
\caption{Real (solid black) and imaginary (dashed) parts of the solution~\eref{isosol} inside of the BH's horizon. As in the previous plot, the thin lines locate the positions of the horizon in each of the coordinates, and the BH has unity mass.}\label{isoellipse}
\end{figure}

The imaginary part of~\eref{isosol} is half an ellipse (positive or negative) enclosed between the values of the horizon in isotropic and Schwarzschild coordinates. It is shown in dashed lines in~\fref{isoellipse} and is expressed parametrically as $\left\{\rthor\frac{1-\cos(\pi x)}{2},\rbhor\sin(\pi x)\right\}$. The real part is just $\frac{1}{2}\left(\rtilde-\mbh\right)$ for the Schwarzschild BH, also shown in the figure. In the superposed case, the real part of the solution is more complicated, which is why~\eref{complexrbar} includes other quantities as factors, and does not reduce to the BH expression just stated. 

\section{Construction of conformal diagrams}\label{diagrams}

Conformal diagrams, also called Carter-Penrose or simply Penrose diagrams, are a useful way to illustrate the causal structure of a spacetime. Created via conformal methods \cite{PhysRevLett.10.66}, they display the complete spacetime compactified along the null directions. Among the extensive literature on conformal diagrams, let us highlight \cite{Vano-Vinuales:2023pum}, because it uses the same infrastructure as here and shows in detail how to construct Penrose diagrams for hyperboloidal slices of Schwarzschild or collapsing spacetimes.

In general, the causal part of the line element of the spacetime to be plotted as a conformal diagram has to be written in the form 
\begin{equation}\label{lincaupen}
d\tilde s^2 = \Xi^2\left(-d\ttilde^2+d\rttort^2\right) \equiv -\Xi^2\, d\tilde u\, d\tilde v , 
\end{equation}
where $\rttort$ would be the tortoise coordinate in case of a BH spacetime, but here it is used in a broader sense for any spherically symmetric spacetime. The null coordinates are $\tilde u = \ttilde-\rttort$ and $\tilde v = \ttilde + \rttort$, and $\Xi$ is the conformal part of the metric. Relevant for the construction of the diagram is how the original metric is put into the form of~\eref{lincaupen}. 

\subsection{Single neutron star}

To express the $(\rtilde,\ttilde)$ part of~\eref{metricuncomp} as~\eref{lincaupen}, 
\begin{equation}\label{nscaupen}
d\tilde s^2 = e^{\nu(\rtilde)}\left(-d\ttilde^2+\frac{e^{\lambda(\rtilde)}}{e^{\nu(\rtilde)}}d\rtilde^2\right) \equiv e^{\nu(\rtilde)}\left(-d\ttilde^2+d\rttort^2\right),
\end{equation}
the relation to be satisfied between the tortoise-like coordinate $\rttort$ and $\rtilde$ is
\begin{equation}\label{srtort}
d\rttort = \sqrt{\frac{e^{\lambda(\rtilde)}}{e^{\nu(\rtilde)}}}d\rtilde . 
\end{equation}
For the spacetime outside of the NS's surface, the square root above equals $\left(1-\frac{2\mns}{\rtilde}\right)^{-1}$, so that the solution to~\eref{srtort} integrates analytically to
\begin{equation}\label{srtort}
\rttort = \rtilde +2\mns\log(\rtilde-2\mns) + C_*, 
\end{equation}
where $C_*$ is an integration constant that will be used to ensure continuity of $\rttort$ at the star's surface. Inside, \eref{srtort} is integrated numerically (using $\nu(\rtilde)$ and $\lambda(\rtilde)$ found in \ssref{cauchytov}) between $\rtilde\in[0,\rtsurf]$, satisfying $\rttort(\rtilde=0)=0$. For the parameter values considered, $C_*=0.7143419$ makes $\rttort(\rtilde)$ continuous, although the matching point at the NS's surface is not smooth. This is a property of the spacetime constructed with the chosen parameters. 

The remaining ingredient to create the conformal diagram is the compactification of the metric in~\eref{nscaupen} along the null coordinates, for which the common choice of the $\arctan$ is made: 
\begin{subequations}\label{basictrafo}
\begin{eqnarray}
 \tilde U = \tilde t - \rttort, \ && \tilde V = \tilde t+\rttort, \label{basic1} \\
 U=\arctan\tilde U, \ && V=\arctan\tilde V, \label{basic2} \\
 T = \frac{V+U}{2}, \ && R = \frac{V-U}{2}. \label{basic3}
\end{eqnarray}
\end{subequations}
Plotting $T$ in the vertical axis and $R$ in the horizontal one (within their ranges of $[-\pi/2,\pi/2]$) for specific constant values of $\ttilde$ and for $\rtilde\in(0,\infty)$ gives the blue curves in the left diagram in \fref{penns}. They look indistinguishable from the equivalent construction of the Minkowski spacetime (in black), whose construction is described in detail in~\cite{Vano-Vinuales:2023pum}. See the same reference for expressing $R$ and $T$ in terms of the compactified radial coordinate $r$. 

\begin{figure}[h]
\includegraphics[width=0.95\linewidth]{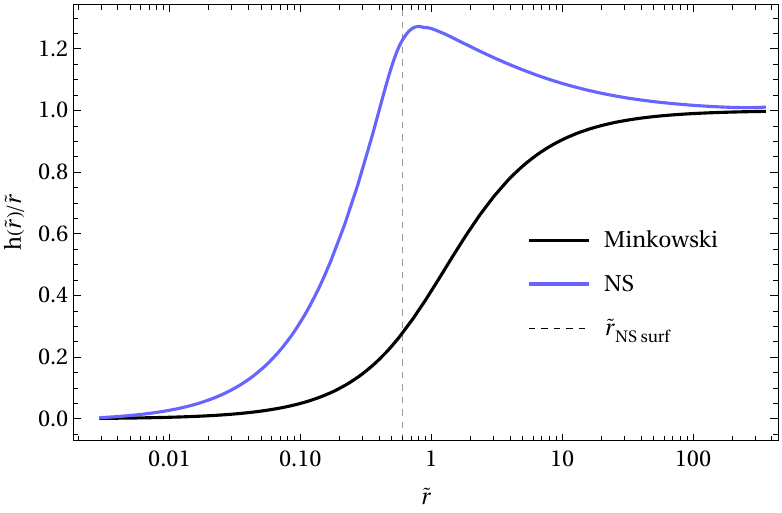}
\vspace{-2ex}
\caption{Height function over $\rtilde$ as a function of $\rtilde$ in logarithmic scale for the NS in blue, and equivalent for the Minkowski case with $h(\rtilde)=\sqrt{\left(\frac{3}{\Kc}\right)^2+\rtilde^2}+\frac{3}{\Kc}$ in black.}\label{heightoverrns}
\end{figure}

For displaying the hyperboloidal slices of the NS spacetime on the right in~\fref{penns}, the integrated height function in~\eref{ttrafo} is needed to transform to the hyperboloidal time $t$, so that the boost function in~\eref{boostexprin} (in~\eref{boostexprout} for outside of the NS surface) needs to be integrated numerically. The result is shown in~\fref{heightoverrns}, where for convenience of visualization $h(\rtilde)/\rtilde$ is plotted.

\subsection{Constraint-satisfying superposition of NS+BH}

\begin{figure}[h]
\includegraphics[width=0.95\linewidth]{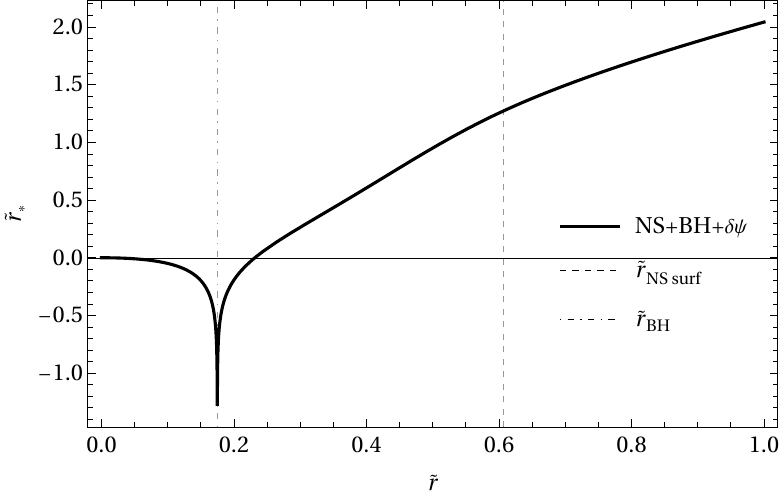}
\vspace{-2ex}
\caption{Tortoise-like coordinate as a function of the areal one for the superposed NS and BH with correction.}\label{rtortvsrboth}
\end{figure}

The choice of $\ttilde\ttilde$ component indicated by~\eref{gttchoice} for the metric describing the superposition is such that the equivalent to~\eref{srtort} takes the form 
\begin{equation}
d\rttort = \frac{1}{\Aa(\rtilde)}d\rtilde . 
\end{equation}
The numerical $\Aa(\rtilde)$ obtained in~\ssref{ssareal} is first integrated between the origin and the NS's surface. The consequence of $\Aa(\rthor)=0$ is that the tortoise-like coordinate will diverge at the BH's horizon. The conditions imposed in the integration are $\rttort(0)=0$ and $\rttort(\rthor^+)=\rttort(\rthor^-)$, where the $\pm$ superscripts denote approaching the horizon from larger ($+$) or smaller ($-$) radii. 
Outside of the NS's surface, $\Aa(\rtilde)=1-\frac{2M}{\rtilde}$ with $M$ the total mass of the system given by~\eref{totalmass}. The expression of the tortoise-like coordinate is given by~\eref{srtort} with $\mns$ substituted by $M$ and, for the chosen parameters, $C_*=0.967968$ makes it continuous at the star's surface. The result is shown in~\fref{rtortvsrboth}. 

For a spacetime including a BH (like the present one), it makes sense to express the metric in coordinates that are well-behaved everywhere except for the singularity. These are Kruskal-Szekeres-like coordinates and they include a compactification along the null directions that yields 
\begin{equation}\label{ed:TR}
\tilde T=e^{\frac{\tort}{B}}\sinh\left(\frac{\tilde t }{B}\right) , \ \tilde R_*=e^{\frac{\tort}{B}}\cosh\left(\frac{\tilde t }{B}\right) 
\end{equation}
for the outer spacetime, and exchanging $\cosh$ by $\sinh$ and vice versa for the region inside of the horizon. 
This is a textbook transformation that is given e.g.~from~(4) in~\cite{Vano-Vinuales:2023pum} (where $\tilde R$ is used instead of $\tilde R_*$). The choice $B=4\,m_{\textrm{\tiny BH superposition}}$ make the metric be finite at the horizon. Note that $m_{\textrm{\tiny BH superposition}} =\msm(\rtilde=10^{-6}) > \mbh=\mns/3$. $\tilde T$ and $\tilde R_*$ are to be substituted into~\eref{basic1} to construct the $T$ and $R$ for the diagram. 

Integration of the boost function into the height function takes place in three different regions: inside of the horizon $\rtilde\in(\rtilde\trum,\rthor)$, not reaching for radii smaller than the location of the trumpet; between the horizon and the NS's surface $\rtilde\in(\rthor,\rtsurf]$; and outside of the star $\rtilde\in[\rtsurf,\infty)$, with $\infty$ approximated by a large number ($O(10)$ is enough in this case). As the leading order behaviour of the compactification factor found from~\eref{confflatnessboth} is $\aconf|_{r=0}\sim r/\rtilde\trum$, the location of the trumpet can be calculated from $\rtilde\trum=\left(\aconf'|_{r=0}\right)^{-1}$. The height function diverges at the trumpet and the horizon. At the latter again the integration conditions $h(\rthor^+)=h(\rthor^-)$, whereas continuity of $h$ is imposed at the surface $\rtsurf$. Substituting $\ttilde$ by $t+h$ in~\eref{ed:TR} will provide hyperboloidal slices of the spacetime with the NS and BH superposed for constant values of $t$, as shown in~\fref{penboth}. 

{
\section{Hyperboloidal initial data for BSSN/Z4 variables}\label{inidata}
The BSSN/Z4 \cite{NOK,PhysRevD.52.5428,Baumgarte:1998te,Brown:2009dd,bona-2003-67,Bernuzzi:2009ex} formulations of the Einstein equations rely on a 3+1 decomposition where the line element usually takes the form
\begin{equation}
    ds^2 = -\left(\alpha^2-\beta^i\beta_i\right) dt^2 + \frac{1}{\chi}\left(2\gamma_{ij}\beta^idx^j+\gamma_{ij}dx^idx^j\right)\,. 
    \label{lineelement}
\end{equation} 
In free evolution, the evolved quantities are the lapse $\alpha$, the shift $\beta^i$, the spatial conformal factor $\chi$ and the conformally rescaled spatial metric $\gamma_{ij}$. The extrinsic curvature tensor $\bar{K}_{ij}=-\frac{1}{2\alpha}\left(\partial_t-\mathcal{L}_\beta\right)\frac{\gamma_{ij}}{\chi}$ is divided into its trace and trace-free parts: 
\begin{subequations}
\begin{eqnarray}
\bar{K} &=& \chi{\gamma}^{ij}\bar{K}_{ij},\\
A_{ij} &=& \chi \bar{K}_{ij} - \frac{1}{3}\gamma_{ij}\bar{K}\,.
\end{eqnarray}
\end{subequations}
The trace $\bar{K}$ is mixed with $\Theta$, a scalar belonging to the Z4 part of the formulation: 
\begin{equation}
    K = \bar{K} - 2 \Theta\, .
    \label{eq:kdefinition}
\end{equation}
Lastly and mixing with the Z4 quantity $Z^i$,
\begin{equation}
    \Lambda^i = \Delta \Gamma^i + 2 \gamma^{ij}Z_j\, 
\end{equation}
is introduced to make the system of equations strongly hyperbolic. Here $\Delta \Gamma^i = \gamma^{jk}\left(\Gamma^i_{jk} - \hat{\Gamma}^i_{jk}\right)$, with $\hat{\Gamma}_{jk}^i$ the connection constructed from a time-independent spatial metric $\hat\gamma_{ij}$, set for convenience to the flat spatial metric. 
}

{
The initial data developed here are expressed on compactified hyperboloidal slices. A suitable method for treating this data is conformal compactification \cite{PhysRevLett.10.66,Friedrich:2003fq,Frauendiener2004,Vano-Vinuales:2014koa,Vano-Vinuales:2017qij,Vano-Vinuales:2024tat}. Its effect is to counteract the blowup of the compactified physical metric $\tilde g_{ab}$ with a rescaling of the form $g_{ab}=\Omega^2\tilde g_{ab}$,
where the rescaled metric $g_{ab}$ is finite everywhere and the conformal factor $\Omega$ vanishes at infinity (set at $r=1$ in this work). The latter's explicit form is taken to be 
\begin{equation}
\Omega=(-\Kc)\frac{1-r^2}{6}\,,
\end{equation}
which satisfies that a CMC slice of Minkowski is written in explicitly flat form and coincides with the solid black lines in figures~\ref{compactns} and ~\ref{aconfboth}. The hyperboloidally compactified initial data for the NS case (\sref{ns}) is conformally rescaled in~\eref{metricnshypcompconf}, while the NS with central BH construction (\sref{nsbh}) is to be conformally compactified as in~\eref{confresclinenemboth}. Comparing those line elements to~\eref{lineelement} above allows to read off the data for each of the metric components of the conformally rescaled metric $g_{ab}$, and use those to construct the remaining evolution variables. 
}

{
In spherical symmetry the evolution variables are $\gamma_{rr}$, $\chi$, $\alpha$, $\beta^r$, $A_{rr}$, $\Lambda^r$, $\Delta\tilde{K}$, and $\tilde{\Theta}$, where the two last ones are defined as
\begin{equation}
    \Delta\tilde{K} = \Omega K - \frac{3 \beta^i \partial_i \Omega}{\alpha} - \Kc\,,\quad \tilde{\Theta} = \Omega \Theta\,.
    \label{eq:ktransformations}
\end{equation} 
The choice of CMC slices implies that the initial data for $\Delta\tilde{K}=0$, and $\tilde\Theta=0$ needs to hold for the initial data to be a solution of the Einstein equations. 
The compactification factor $\aconf$ was determined in a way -- solving~\eref{confflatnessns} and~\eref{confflatnessboth} -- such that the spatially rescaled metric $\gamma_{ij}$ is flat, implying that $\gamma_{rr}=1$ and $\Lambda^r=0$. 
The profiles of the remaining evolution variables for each of the scenarios considered are shown in~\fref{vars}. As expected, the lapse and the spatial conformal factor $\chi$ are zero at the origin (the location of the puncture) in the presence of a BH, and the shift is positive in the area around the horizon. The $A_{rr}$ component of the trace-free part of the extrinsic curvature is zero at the origin in the Minkoswki and single NS cases, while it takes a negative finite value at the BH's trumpet location (puncture at $r=0$ in the BH scenarios). Note that the profiles are regular for all values of the radial coordinate, including the origin / puncture location and \scrip, so that these data can be straightforwardly interpolated onto the desired radial grid and evolved in a spherically symmetric code. 
\begin{figure}[h]
\includegraphics[width=0.95\linewidth]{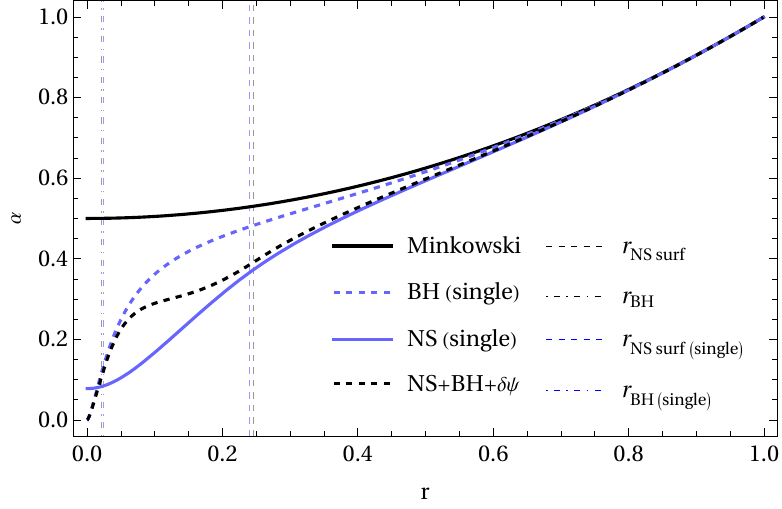}
\\
\includegraphics[width=0.95\linewidth]{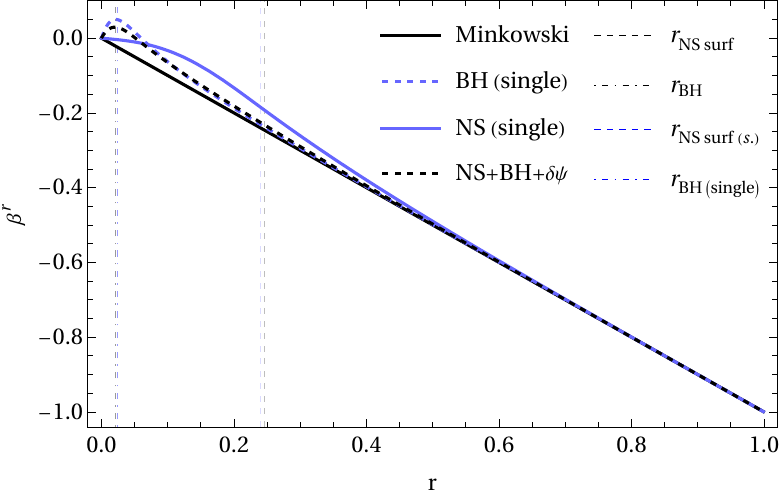}
\\
\includegraphics[width=0.95\linewidth]{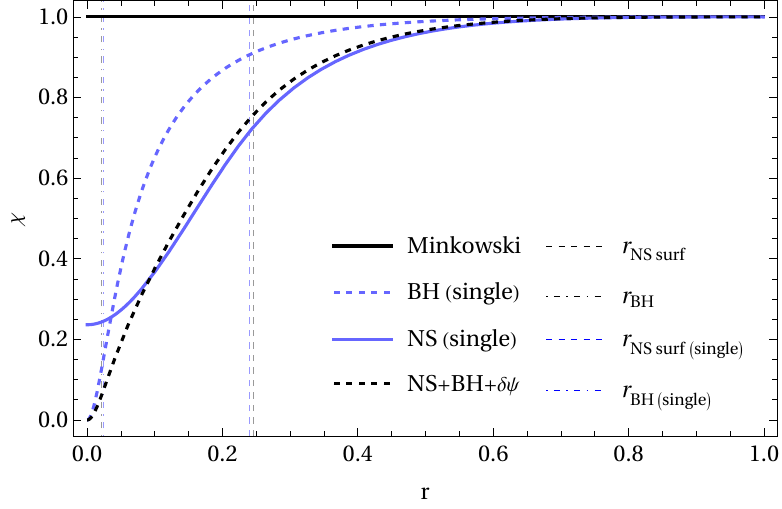}
\\
\includegraphics[width=0.95\linewidth]{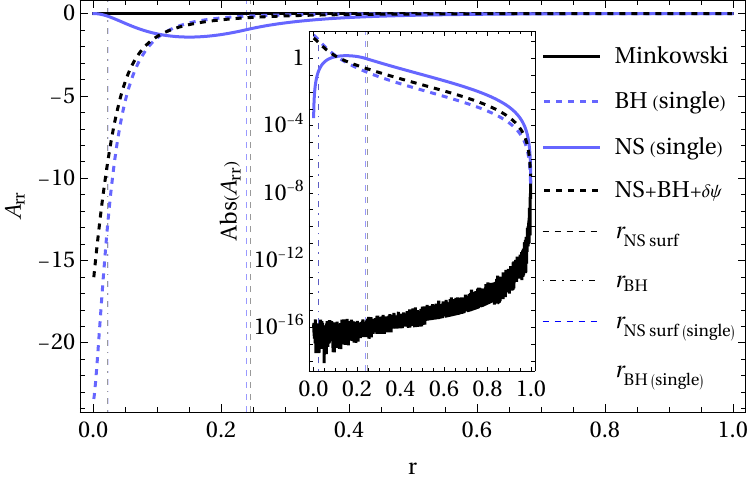}
\vspace{-2ex}
\caption{{Initial data for (from top to bottom) the conformally rescaled lapse $\alpha$, radial component of the shift $\beta^r$, spatial conformal factor $\chi$ and $rr$ component of the trace-free part of the conformal extrinsic curvature $A_{rr}$ for the different scenarios considered in this work. The inset in the latter shows more clearly the differences between the Minkowski case and those with central objects. The legend is the same as in~\fref{aconfboth}.}}\label{vars} 
\end{figure}
}
\bibliography{hypcomp}

\end{document}